\documentclass[12pt]{article}

\AtBeginDocument{
  \addtocontents{toc}{\normalsize}%\footnotesize
}
\pdfoutput=1

\usepackage{amsmath,amssymb,amsfonts,amsthm,amscd,mathrsfs}
\usepackage{xcolor}
\definecolor{darkblue}{rgb}{0.1,0.1,.7}
\usepackage[]{version}
\usepackage{float}
\usepackage[]{graphicx}
\usepackage[]{latexsym}
\usepackage{geometry}

\usepackage{subcaption}
\usepackage[utf8]{inputenc}
\usepackage[all,cmtip]{xy}
\usepackage{empheq}

\usepackage{ifthen}
\usepackage{tikz}
\usepackage{array,setspace,yfonts,dsfont,bbm,colonequals}
\usepackage[colorlinks, linkcolor=darkblue, citecolor=darkblue, urlcolor=darkblue, linktocpage]{hyperref} 

\usepackage{cite}
\usepackage{xspace}
\usepackage{empheq}
\usepackage{xcolor}

\usepackage{caption}
\usepackage{subcaption}
\usepackage{booktabs}
\usepackage{siunitx}
\numberwithin{equation}{section}
\newcommand{\ba}{\begin{equation}\begin{aligned}}
\newcommand{\ea}{\end{aligned}\end{equation}}

\newcommand{\Ll}{\underline \Lambda_{w,\bar w}}
\newcommand{\Lu}{\overline \Lambda_{w,\bar w}}

\newcommand{\cO}{\mathcal O}

\newcommand{\reef}[1]{(\ref{#1})}
\newcommand{\be}{\begin{equation}}
\newcommand{\ee}{\end{equation}}

\newcommand{\bea}{\begin{equation} \begin{aligned}}
\newcommand{\eea}{\end{aligned}\end{equation}}

\newcommand{\ud}{\mathrm d}

\newcommand{\Gm}{\mathcal G_{\mbox{\tiny inf}}}
\newcommand{\GM}{\mathcal G_{\mbox{\tiny sup}}}

\newcommand{\Df}{{\Delta_\phi}}

\begin{document}

\vspace*{-.6in} \thispagestyle{empty}
\begin{flushright}
%CERN PH-TH/2015-200\\
%LPTENS/18/18
\end{flushright}
%%
%% Title
%%
\vspace{1cm} {\Large
\begin{center}
{\bf Bounding 3d CFT correlators}\\
\end{center}}
\vspace{1cm}
%%%
%% Authors
%%%
\begin{center}
{Miguel F.~Paulos and Zechuan Zheng}\\[1cm] 
{
\small
{\em Laboratoire de Physique de l'\'Ecole Normale Sup\'erieure\\ PSL University, CNRS, Sorbonne Universit\'es, UPMC Univ. Paris 06\\ 24 rue Lhomond, 75231 Paris Cedex 05, France}
}\normalsize
\\
%\vspace{1cm}\today
\end{center}

\begin{center}
	{\texttt{miguel.paulos@ens.fr},
	\texttt{zechuan.zheng@ens.fr}
	}
	\\
	%\vspace{1cm}\today
\end{center}

\vspace{4mm}

\begin{abstract}
We consider the problem of bounding CFT correlators on the Euclidean section. By reformulating the question as an optimization problem, we construct functionals numerically which determine upper and lower bounds on correlators under several circumstances. A useful outcome of our analysis is that the gap maximization bootstrap problem can be reproduced by a numerically easier optimization problem.
We find that the 3d Ising spin correlator takes the minimal possible allowed values on the Euclidean section.  Turning to the maximization problem we find that for $d>2$ there are gap-independent maximal bounds on CFT correlators. Under certain conditions we show that the maximizing correlator is given by the generalized free boson for general Euclidean kinematics. In our explorations we also uncover an intriguing 3d CFT which saturates gap, OPE maximization and correlator value bounds. Finally we comment on the relation between our functionals and the Polyakov bootstrap. 
\end{abstract}
\vspace{2in}

%\hspace{0.7cm} March 2018

\newpage

{
\setlength{\parskip}{0.05in}
\tableofcontents
\renewcommand{\baselinestretch}{1.0}\normalsize
}

%\newpage

\setlength{\parskip}{0.1in}
\newpage

\section{Introduction}\label{sec:introduction}

\begin{figure}
	\begin{center}
		\includegraphics[width=10cm]{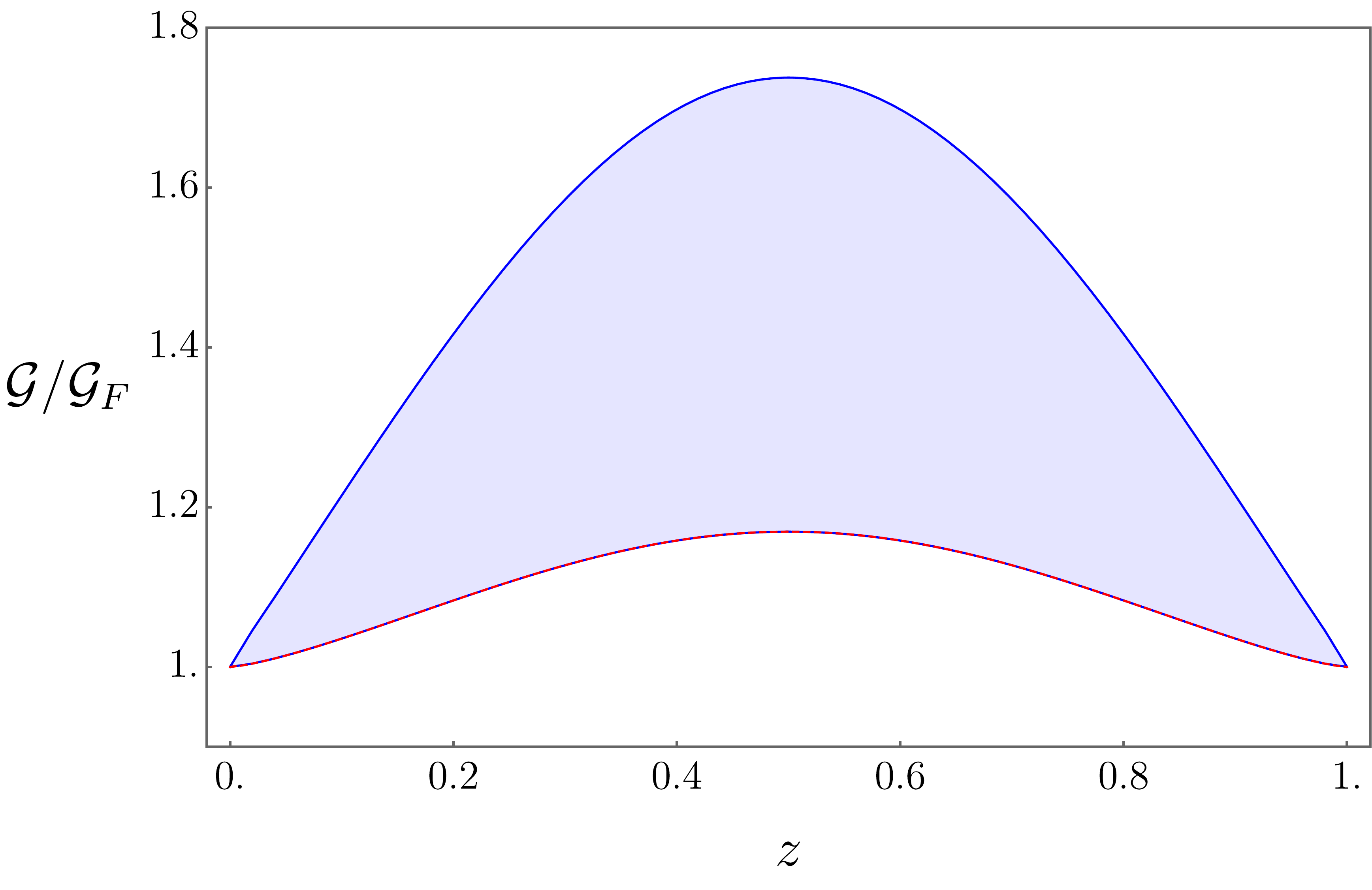}
		\caption{Upper and lower bounds on 3d CFT correlators  $\mathcal G(z,z)$ normalized by $\mathcal G^F(z)=-1+z^{-2\Df}+(1-z)^{-2\Df}$. Here $\Df=\Delta_{\sigma}^{\mbox{\tiny Ising}}\sim 0.518149$. The 3d Ising spin four-point function saturates the lower bound.}
	\end{center}
\end{figure}

The last few years have taught us that CFTs and in particular the CFT data which defines correlators of local operators, such as quantum numbers and OPE coefficients, are strongly constrained \cite{Poland:2018epd}. These constraints imply that the CFT theory space is bounded along certain directions. More remarkably, interesting theories tend to lie at the edges of the realms of possibility \cite{El-Showk:2012cjh}, realms ruled by the iron fists of crossing symmetry and unitarity.

Up to now, the collective efforts of the conformal bootstrap community have explored this space by focusing on the constraints of a restricted set of four-point correlators and seeing how far one can move along two directions. In the first direction, one determines an upper bound on the scaling dimension of the first operator appearing in the OPE with certain quantum numbers - the gap maximization problem \cite{Rattazzi:2008pe}. In the second one determines upper (or lower) bounds on the OPE coefficient of some particular operator appearing in these correlators \cite{Rattazzi:2010gj}. The bounds are determined by constructing appropriate functionals, numerically or analytically.
These two directions are like two different viewpoints on the CFT landscape. 

In this paper we will introduce new ones. The idea is simple: we will determine bounds on the values of CFT correlators. This can be thought of as constraining very special infinite linear combinations of OPE coefficients. This is a very natural question which surprisingly has not been considered so far. An important motivation is the hope that with this new lens we may reach interesting and yet unseen corners of CFT space. Indeed, it is known that the analogous problem of minimizing or maximizing values of $S$-matrices is known to lead to bounds which can be saturated by interesting theories. This is in fact more than an analogy, since we now know that $S$-matrices are intimately related to (limits of) CFT correlators \cite{Paulos:2016fap}. We will see that the corresponding CFT problem is equally interesting.

We consider CFT four-point correlators with spacelike separated operators, so that the correlator is real, and we will obtain exact and numerical bounds on its values. The exact bounds are simple consequences from recent work \cite{Paulos:2020zxx}, where one of us obtained rigorous correlator bounds for the special case where these operators all lie on a line (or a circle). These bounds state that
\ba
\mathcal G(w,w)&\geq \mathcal G^F(w) \\
\mathcal G(w,w)&\leq \mathcal G^B(w)\,, \qquad \Delta_g\geq 2\Df \\
\ea
where $\mathcal G^{B,F}(w):=\pm 1+w^{-2\Df}+(1-w)^{-2\Df}$ have the interpretation of generalized free field (GFF) correlators, and the condition on the gap $\Delta_g$ is imposed across all spins. The upper bound is easily generalisable to more general Euclidean configurations. More interestingly, we can obtain bounds in Lorentzian kinematics where $w,\bar w$ are real and independent. In particular we will obtain that the correlator is parametrically bounded in the double lightcone limit:
\bea
\mathcal G(w,1-w)\underset{w\to 0^+}{\gtrsim} &\frac{1}{w^{\Df}}\\
\mathcal G(w,1-w)\underset{w\to 0^+}{\lesssim}&  \frac{1}{w^{3\Df}}\,.
\eea

We can obtain numerical bounds by constructing appropriate linear functionals that act on the crossing equation for the correlator and satisfy appropriate positivity properties. While such bounds can be obtained for any spacelike configuration, in practice lorentzian configurations are highly sensitive to the large spin spectrum and so are difficult to study numerically. We hence focus on Euclidean configurations in this work. We will determine both upper and lower bounds for CFTs in $d=3$. Notice that the exact lower bound above is interesting as it is saturated by the same correlator which maximizes the gap in $d=1$. This is not an accident, and in fact we will see that under certain circumstances, the same is true in higher dimensions. In particular we will establish, numerically:
\bea
\mathcal G(w,\bar w)\geq \mathcal G_{\mbox{\tiny gapmax}}(w,\bar w)\,, \qquad \Df\lesssim d/2
\eea
where $\mathcal G_{\mbox{\tiny gapmax}}$ stands for the correlator which maximizes the scalar gap. In particular, using the fact the the 3d Ising spin correlator maximizes the gap, we get
\bea
\mathcal G(w,\bar w)\geq \mathcal G_{\mbox{\tiny Ising}}(w,\bar w)\,, \qquad \Df=\Delta_{\sigma}^{\mbox{\tiny Ising}}\sim 0.518149
\eea
for any 3d CFT correlator with that $\Df$. An interesting outcome of this equivalence is that we can solve the gap maximization problem in terms of one of correlator minimization\footnote{In fact more generally it is easier to define a simpler, related problem, see section \ref{sec:opt}.}. This has the advantage that solving the latter problem requires a single optimization step, whereas the former requires a costly binary search. In this way for instance we check that we can get in a single step the correct maximal gap to within $10^{-15}$.

The maximization problem is also interesting. In $d=3$ there is no need to impose a gap in the spectrum to obtain an upper bound on the correlator if $\Df\leq d-2$. We find that this upper bound is saturated by a different CFT depending on the point at which we maximize, thereby giving us an entire family of bound saturating solutions to crossing, a family that generically contains a stress-tensor. By maximizing the correlator in the OPE limit we find a family of theories containing a protected operator of dimension $5/2$. Remarkably, this family coincides with that obtained by maximizing the gap at a special point $\Df\sim 0.505$.

If we do impose a gap, we find that the maximal correlator coincides, in a wide regime of parameters, with the correlator which maximizes the OPE coefficient of the operator at the gap. For the special value $\Delta_g=2\Df$ our numerical bound must match with the exact bound above, at least for 1d kinematics, since it is saturated by the generalized free boson solution which exists in any dimension. We determine that this remains unchanged in the entire Euclidean section. We also show that nearby deformations of this solution, computed by certain AdS contact interactions, also saturate this bound to leading order.

The outline of this work is as follows. Section 2 is concerned with describing general kinematics of CFT correlators as well as giving an overview of the space of such correlators and some notable points in that space. In section \ref{sec:expect} we make some simple observations regarding what to expect for the problem of maximizing and minimizing CFT correlator values for spacelike separated operators. In particular, we show how exact results recently derived for correlators on a line can be used to obtain bounds for more general kinematics. In section \ref{sec:opt} we show how linear optimization methods based on linear functionals can be used effectively to place both lower and upper bounds on correlators. In the process we show that the bootstrap question of maximizing a gap can be reformulated as a certain correlator minimization problem. The upshot is that this can be solved in a single optimization step. Section \ref{sec:numapp} is concerned with numerical applications and contains our main results, deriving both upper and lower bounds on 3d CFT correlators by numerically constructing appropriate functionals. We finish this paper with a short discussion and an appendix containing details of our numerical implementations.

\section{Conformal correlators}

\subsection{Kinematics and crossing}
We interested in CFT four-point correlators of identical scalar operators, which we write as as
\bea
\langle \phi(x_1)\phi(x_2)\phi(x_3)\phi(x_4)\rangle=\frac{\mathcal G(u,v)}{x_{13}^{2\Df} x_{24}^{2\Df}}\,, 
\eea
with $x_{ij}^2\equiv (x_i-x_j)^2$. The correlator depends on the two conformal cross-ratios $u,v$ which can be written in terms of two complex variables $z, \bar z$, determined as
\bea
u\equiv z \bar z\equiv \frac{x_{12}^{2} x_{34}^2}{x_{13}^2 x_{24}^2}\,,\qquad v\equiv (1-z)(1-\bar z)\equiv \frac{x_{14}^{2} x_{23}^2}{x_{13}^2 x_{24}^2}\,.
\eea
It will be also useful to consider yet a different set of coordinates, $\rho,\bar \rho$, defined as \cite{Hogervorst:2013sma}:
\bea
\rho_z:= \frac{1-\sqrt{1-z}}{1+\sqrt{1-z}}\,,\quad z=\frac{4\rho_z}{(1+\rho_z)^2}, \qquad \mbox{abbreviated as}\quad \rho\equiv \rho_z, \bar \rho\equiv \rho_{\bar z}\,,
\eea
We set $\mathcal G_z(z,\bar z)\equiv \mathcal G(u(z,\bar z),v(z,\bar z))$ and $\mathcal G_\rho(\rho,\bar \rho)=\mathcal G_z(z,\bar z)$, but will often abuse notation and drop the subscripts. 

When all operators are spacelike separated, the cross-ratios $u,v$ can range from zero to infinity. This can be split into two regions separated by a third:
\bea
\mbox{Euclidean:}&\qquad (1+u-v)^2>4u \Leftrightarrow z^*=\bar z,\quad z \in \mathbb C\\
\mbox{Lorentzian spacelike:}&\qquad (1+u-v)^2<4u \Leftrightarrow (z,\bar z) \in \mathbb (-\!\infty,0)^2\cup (0,1)^2\cup (1,\infty)^2\\
\mbox{Line:}&\qquad (1+u-v)^2=4u \Leftrightarrow z=\bar z,\quad z\in \mathbb R
\eea
\begin{figure}%
\begin{center}
\begin{tabular}{ccc}
\includegraphics[width=6 cm]{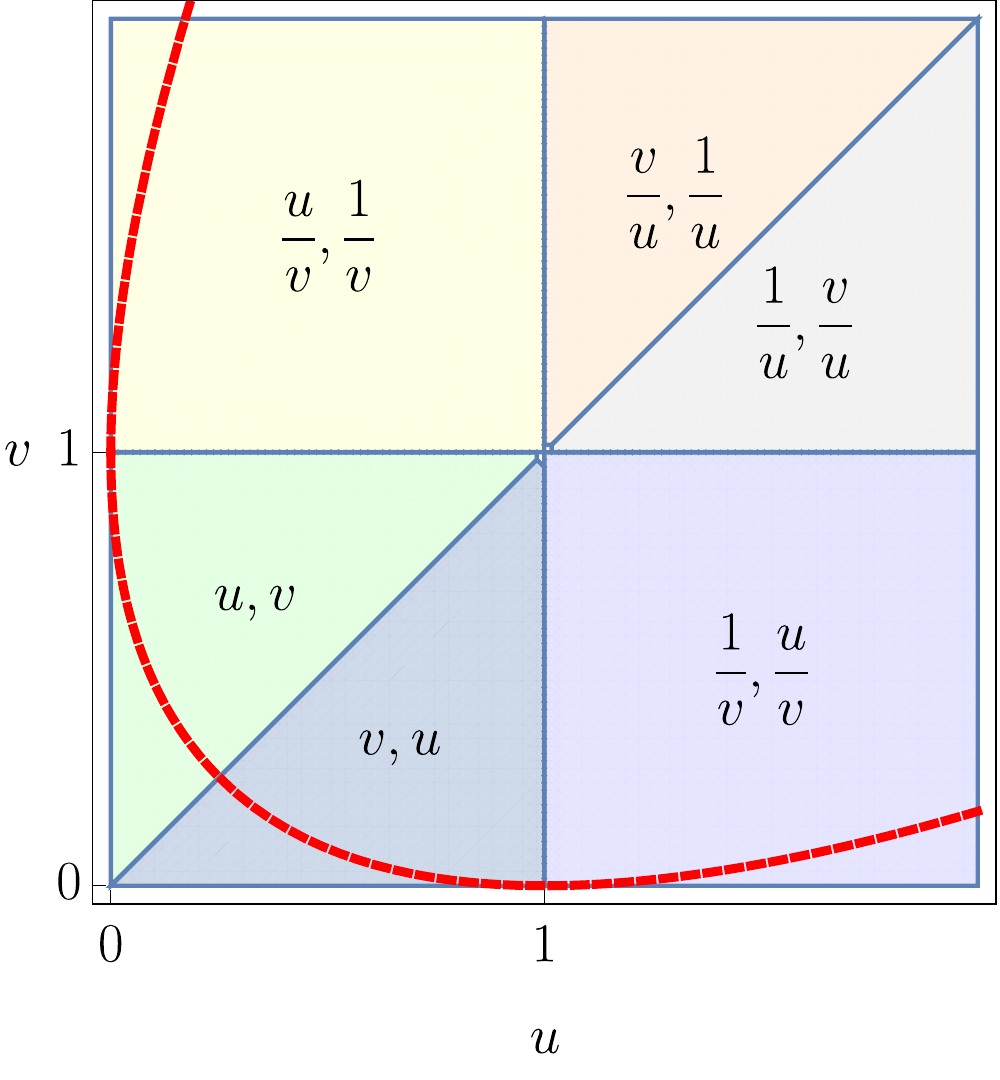}%
&
&
\includegraphics[width=6 cm]{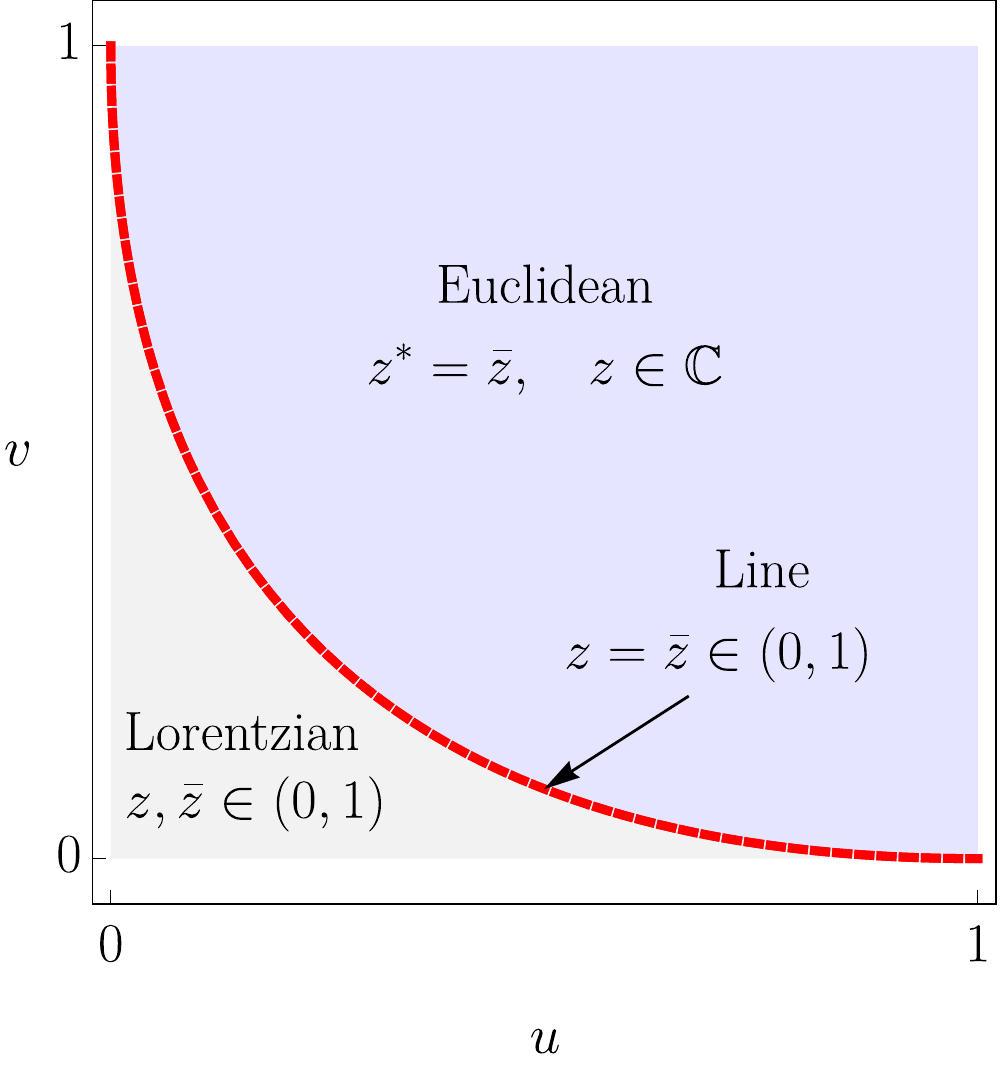}%
\end{tabular}
\caption{Cross-ratio space for spacelike separated operators. The thick red curve corresponds to operators on a line. To its right the Euclidean and to its left the Lorentzian sections. On the lefthand figure we show how this space may be split into regions mapped into each other by crossing transformation. On the righthand figure we zoom in on the region $u,v \in (0,1)$. Our choice of fundamental domain corresponds to taking $v>u$ in that figure.\label{fig:uv}}%%
\end{center}
\end{figure}
Furthermore, Bose symmetry implies the invariances
\bea
\mathcal G(z,\bar z)=\mathcal G(1-z,1-\bar z)=\frac{\mathcal G\left(\frac{1}{z},\frac{1}{\bar z}\right)}{(z\bar z)^{\Df}}\label{eq:crossingrelations}
\eea
which are obtained by swapping $x_1\leftrightarrow x_4$ and $x_1 \leftrightarrow x_3$. By combining these transformations we can also get
\bea
\mathcal G(z,\bar z)=\frac{\mathcal G\left(\frac{z}{z-1},\frac{\bar z}{\bar z-1}\right)}{[(1-z)(1-\bar z)]^{\Df}}=\frac{\mathcal G\left(\frac{1}{1-z},\frac{1}{1-\bar z}\right)}{[(1-z)(1-\bar z)]^{\Df}}=\frac{\mathcal G\left(\frac{z-1}{z},\frac{\bar z-1}{\bar z}\right)}{(z\bar z)^{\Df}}.\label{eq:transf2}
\eea
This means that the spacelike region $u,v>0$ can be divided into six subregions, each of which may be mapped into any other by one of the above transformations. We show this in figure \ref{fig:uv}.

\begin{figure}
\begin{center}
\includegraphics[width=17 cm]{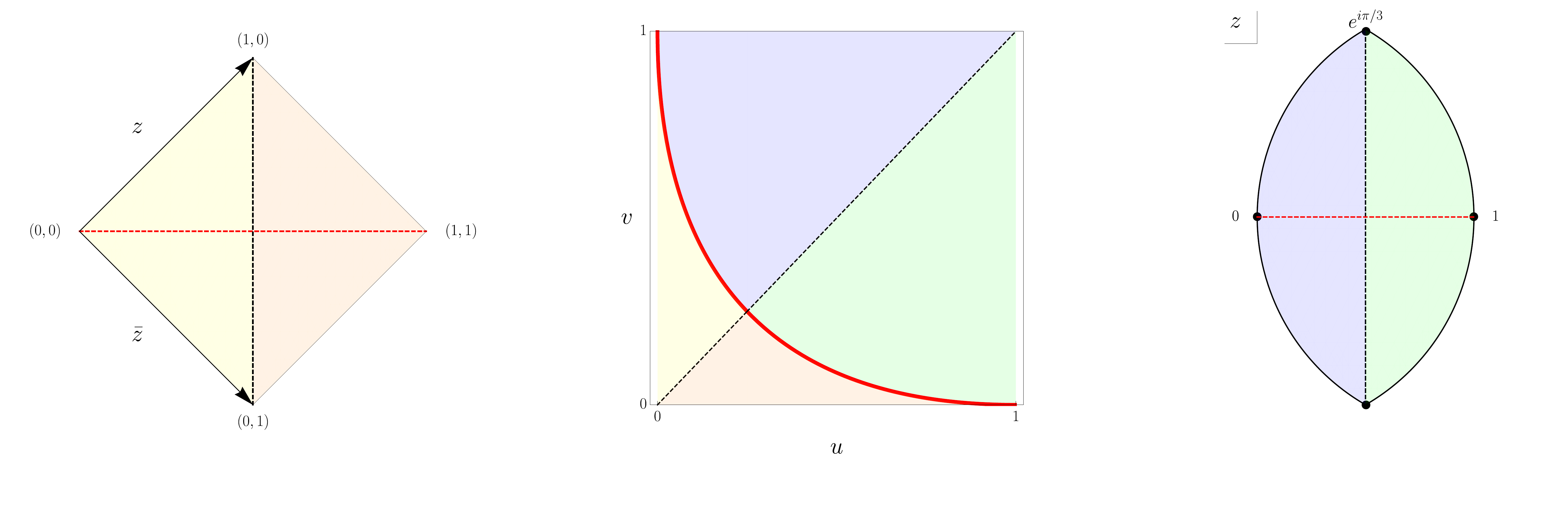}%
\caption{Fundamental domain in cross-ratio space. Our choice of fundamental domain corresponds to taking $v>u$ in the central figure. This captures both a Lorentzian spacelike and an Euclidean sections, shown on the left and right respectively. The arcs on the right lie on circles $|z|=1$ and $|1-z|=1$.}%
\label{fig:funddomain}%
\end{center}
\end{figure}

We will be interested in studying bounds on correlation functions of spacelike operators. The identities above show that it is sufficient to restrict our attention to a particular region in cross-ratio space. We will choose this region to be the set $0<u<v<1$, although it is useful to keep in mind the wider region $0<u,v<1$. We show what this domain corresponds to in terms of the $z$ variables in figure \ref{fig:funddomain}.

The correlator admits an expansion in conformal blocks, which we write as
\bea
\mathcal G(z,\bar z)=\frac{1}{(z\bar z)^{\Df}}+\sum_{\substack{\Delta\geq \Delta_u(\ell)\\\ell=0,2,4\ldots}} a_{\Delta,\ell} \frac{G_{\Delta,\ell}(z,\bar z)}{(z \bar z)^{\Df}}\,.
\eea
The sum ranges over the scaling dimension $\Delta$ and traceless-symmetric spin $\ell$ of exchanged states, which are restricted by
Bose symmetry and unitarity; in particular $\ell$ should be even and $\Delta_u(\ell)=\frac{d-2}{1+\delta_{\ell,0}}+\ell$. We have also separated out the contribution of the identity operator. The coefficients $a_{\Delta,\ell}$ correspond to squares of OPE coefficients between the external operators and the exchanged states, viz.:
\bea
\phi \times \phi \sim \sum_{\cO\in \phi\times \phi} \lambda_{\phi \phi \cO_{\Delta,\ell}} \cO_{\Delta,\ell}\,, \qquad a_{\Delta,\ell}:= \lambda_{\phi \phi \cO_{\Delta,\ell}}^2\,.
\eea
On the Euclidean section the conformal blocks satisfy
\bea
G_{\Delta,\ell}(z,\bar z)=G_{\Delta,\ell}\left(\frac{z}{z-1},\frac{\bar z}{\bar z-1}\right)
\eea
which trivializes some of the identities in \reef{eq:transf2}. The only independent non-trivial constraint is the crossing equation:
\bea
F_{0,0}(z,\bar z|\Df)+\sum_{\substack{\Delta\geq \Delta_u(\ell)\\\ell=0,2,4\ldots}} a_{\Delta,\ell} F_{\Delta,\ell}(z,\bar z|\Df)=0\,,
\eea
where we have defined the crossing vectors
\bea
F_{\Delta,\ell}(z,\bar z|\Df)\equiv \frac{G_{\Delta,\ell}(z,\bar z)}{(z\bar z)^{\Df}}-\frac{G_{\Delta,\ell}(1-z,1-\bar z)}{[(1-z)(1-\bar z)]^{\Df}}\,.
\eea
\subsection{The space of correlators and its special points}
\label{sec:corrspace}
We define the space of all consistent CFT$_d$ four point functions of identical scalar operators with scaling dimension $\Df$:
\bea
\mathfrak G^{(d)}_{\Df}:=\bigg\{\mbox{\em Identical scalar correlators $\mathcal G=\langle 
\phi \phi \phi \phi\rangle$ of CFT$_d$} \bigg\}
\eea
By consistency here, we mean that the correlators satisfy crossing and an OPE expansion consistent with unitarity. In principle we could and should refine this by also demanding that $\mathcal G$ is part of a larger family of correlators involving various external operators and that the full set is consistent with all crossing symmetry constraints. 

Note that since a CFT in $d$ dimensions is also a CFT in any lower dimension, we have the inclusion relations:
\bea
\mathfrak G^{(d')}_{\Df}\subset \mathfrak G^{(d)}_{\Df}\,, \qquad 1<d<d'\,. \label{eq:inclusion1}
\eea
The case $d=1$ is special, since such correlators have more restricted kinematic dependence. In particular they depend on a single cross-ratio. However we can still say that
\bea
\mathcal G\big|_{z=\bar z} \in \mathfrak G^{(d=1)}_{\Df}\qquad \mbox{for any}\quad \mathcal G \in \mathfrak G_{\Df}^{(d)}\,.\label{eq:inclusion2}
\eea
For convenience below we will drop the subscript $\Df$ and the spacetime dimension, but it should be clear all statements below - where we compare values of correlators - refer to fixed choices of these parameters, so that we do not compare apples to oranges.

We will now define several quantities of interest. We begin with the {\em infimum correlator}:
\bea
\Gm(w,\bar w)&:=\mbox{inf}\left\{\mathcal G(w,\bar w)\quad \mbox{for}\quad \mathcal G \in \mathfrak G\right\}\,.
\eea
It is important to emphasize that the infimum correlator is actually not a physical correlator in general i.e. $\Gm$ itself will in general not lie in $\mathfrak G$. The most we can say is that for each choice of $w,\bar w$ there exists a {\em minimizing} or {\em minimal value} correlator at $w,\bar w$, denoted $\mathcal G_{\mbox{\tiny min};w,\bar w}\in \mathfrak G$, such that
$\mathcal G_{\mbox{\tiny min};w,\bar w}(w,\bar w)=\Gm(w, \bar w)$:
\bea
\mathcal G_{\mbox{\tiny min};w,\bar w}= \underset{\mathcal G\in \mathfrak G}{\mbox{arg min}}\  \mathcal G(w,\bar w)
\eea
We can make analogous definitions for maximization. The {\em supremum} correlator is:
\bea
\GM(w,\bar w)&{:=}\mbox{sup}\left\{\mathcal G(w,\bar w)\quad \mbox{for}\quad \mathcal G \in \mathfrak G\right\}
\eea
However there is a catch: it can be the case that the result is infinite. This is essentially due to the fact that in general there are unitary solutions to the crossing equation without the identity operator. Such solutions can always be added with an arbitrarily large coefficient\footnote{An equivalent way of describing the issue is that the OPE maximization problem described below can be unbounded.}, leading to an unbounded size for the correlator. The way out is that such solutions disappear once we set a sufficiently large gap in the spectrum of operator dimensions. Defining $\Delta_g^{\mathcal G}$ to be the lowest non-zero scaling dimension of a scalar operator appearing in the OPE of the correlator $\mathcal G$, we can make the refined definition:
\bea
\GM(w,\bar w|\Delta_g)&:=\mbox{sup}\left\{\mathcal G(w,\bar w)\quad \mbox{for}\quad \mathcal G \in \mathfrak G\quad \mbox{s.t.}\quad \Delta_g^{\mathcal G}\geq \Delta_g\right\}
\eea
As was the case for the minimum value correlator, it is important to emphasize the distinction between the maximal value correlator and the supremum correlators at specific points:
\bea
\mathcal G_{\mbox{\tiny max};w,\bar w}(\bullet|\Delta_g):= \underset{\mathcal G\in \mathfrak G: \Delta_g^{\mathcal G}\geq \Delta_g}{\mbox{arg max}}\  \mathcal G(w,\bar w)
\eea
We could of course refine this further by allowing for different gaps in all the various spin sectors of the spectrum, but this will do for the purposes of this work.

To conclude this section we introduce two further definitions. The first is related to the fact that for a given $\Df$ there is an associated maximal gap which is allowed in any given spin sector. We therefore set:
\bea
\mathcal G_{\mbox{\tiny gapmax}}:= \underset{\mathcal G\in \mathfrak G}{\mbox{arg max}}\quad \Delta_g^{\mathcal G}
\eea

The second definition is similar and related to maximizing the OPE coefficient of the operator with dimension $\Delta_g$ in the set of correlators with a $\Delta^{\mbox{\tiny gap}}_{\mathcal G}=\Delta_g$:
\bea
\mathcal G_{\mbox{\tiny opemax}}(\bullet|\Delta_g):= \underset{\mathcal G\in \mathfrak G:~\Delta_{\mathcal G}^{\mbox{\tiny gap}}\geq \Delta_g}{\mbox{arg max}}\quad a_{\Delta_g,0}\,.
\eea
As we will now discuss, the gapmax and opemax correlators can often be related to the minimal value and maximal value correlators respectively.

\section{General expectations and results}
\label{sec:expect}

\subsection{Basic intuition}
\label{sec:intuition}

\begin{figure}
	\begin{center}
		\begin{tabular}{cc}
			\includegraphics[width=6 cm]{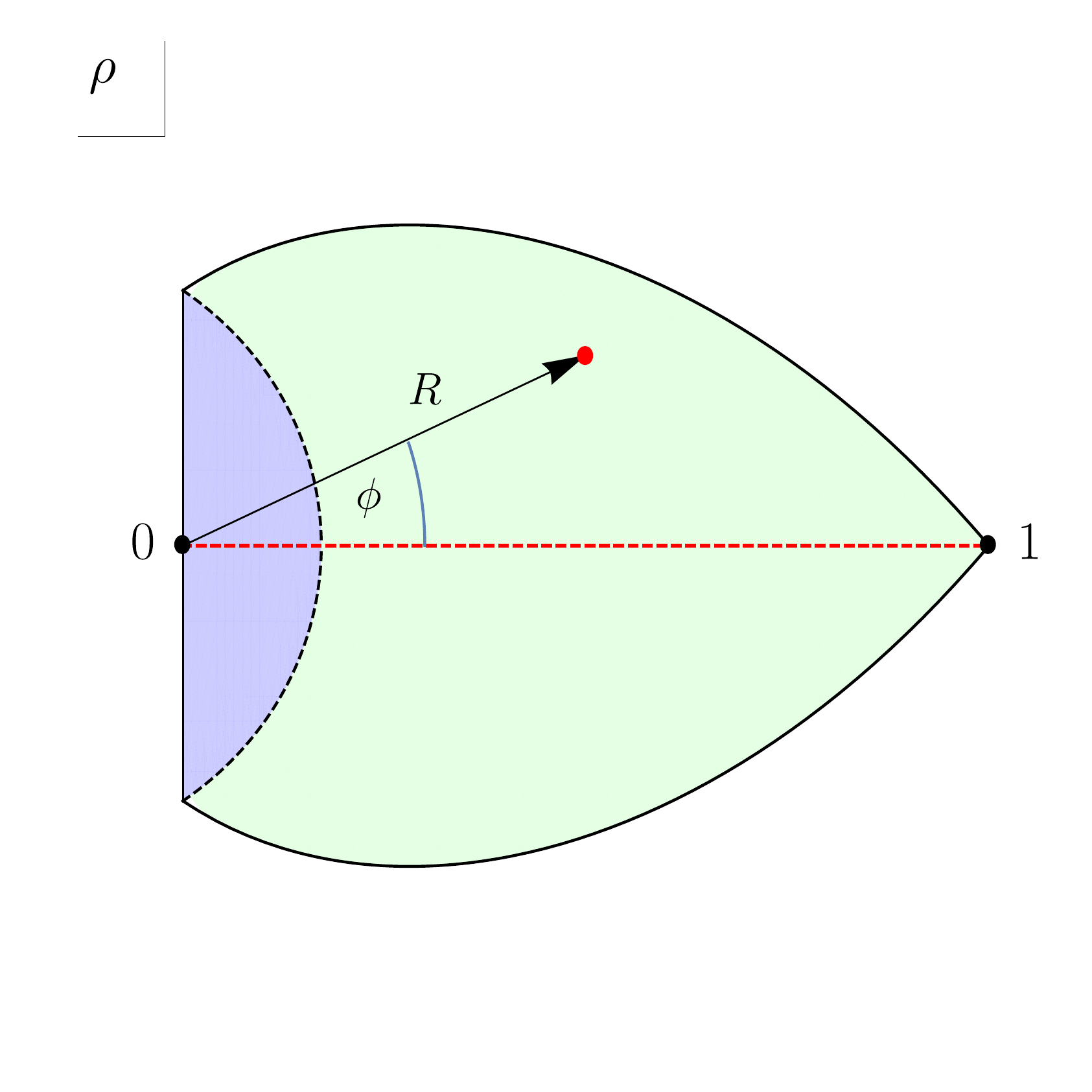}%
			&
			\includegraphics[width=6 cm]{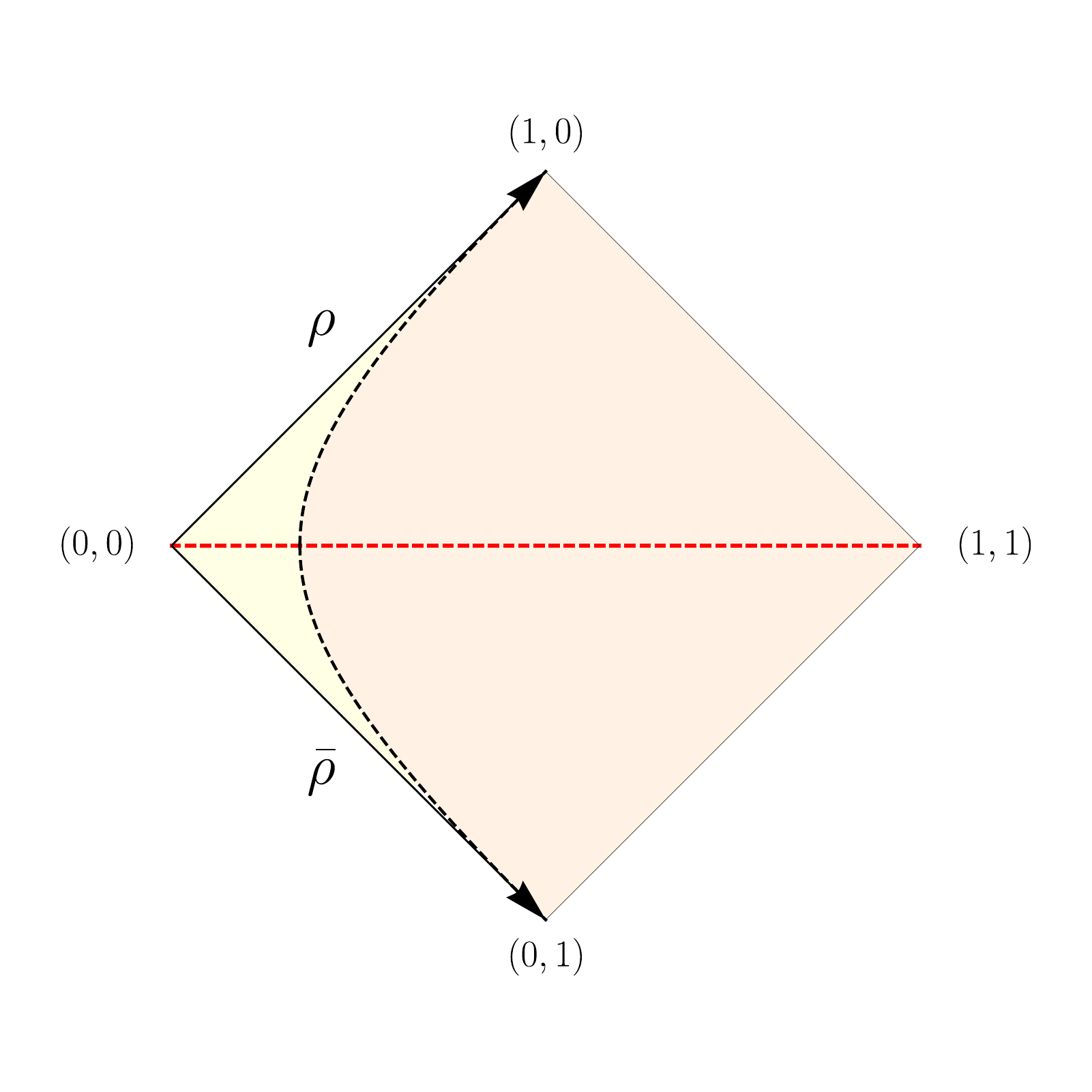}%
		\end{tabular}
		\caption{Fundamental domain in the $\rho$ variable. On the left in the Euclidean section, the fundamental domain is in blue, whereas on the right in the Lorentzian spacelike it is shown in yellow. The two regions shown in each section are mapped into each other by crossing, $z,\bar z\to 1-z,1-\bar z$. Inside the Euclidean fundamental domain we have that $R\leq 3-2\sqrt{2}\sim 0.17 $ if $\phi=0$ and $R\leq 2-\sqrt{3}\sim 0.27$ for $\phi=\pm \pi/2$. In the Lorentzian spacelike section we have $R\leq 3-2\sqrt{2}$, with equality achieved along the line $\rho=\bar \rho$.}%
		\label{fig:funddomainrho}%
	\end{center}
\end{figure}

We would like to get some intuition on what to expect for the bounds on correlator values. For this it is useful to consider the conformal block expansion of the correlator. 
An important property of the conformal blocks is that they not only have a positive expansion in terms of the $z,\bar z$ cross-ratios, but also in terms of $\rho,\bar \rho$ \cite{Pappadopulo:2012jk}. That is:
\bea
G_{\Delta,\ell}(z,\bar z)&=r^{\Delta}\sum_{n=0}^{\infty} \sum_{|j-\ell|\leq 2n} c_{n,j}r^{2n} C^{\left(\frac{d-2}2\right)}_{j}(\cos \theta)\\
&=R^{\Delta}\sum_{n=0}^{\infty} \sum_{|j-\ell|\leq 2n} d_{n,j}R^{2n} C^{\left(\frac{d-2}2\right)}_{j}(\cos \phi)\label{eq:blockexp}
\eea
with $c_{n,j},d_{n,j}\geq 0$ and $z=r e^{i\theta}, \rho=R e^{i\phi}$ are polar representations for $z$ and $\rho$. Notice that in both cases successive terms in the power series are suppressed by the square of the radial variable. This is especially relevant if we work in the $\rho$ variable. As shown in figure \ref{fig:funddomainrho}, in the fundamental Euclidean domain $R$ is never larger than $2-\sqrt{3}\sim 0.268$.
It follows that everywhere inside the Euclidean fundamental domain the conformal blocks are well approximated by keeping only the leading term in the power series representation:
\bea
G_{\Delta,\ell}(\rho,\bar \rho)\propto R^{\Delta} C_\ell^{(\frac{d-2}2)}(\cos\phi)\,.
\eea 

In turn, the smallness of $R$ suggests that the correlator is well approximated by keeping only the lowest dimension non-identity operator in the conformal block expansion. For definiteness we will assume this to be a scalar with dimension $\Delta_g$. Indeed, numerical bootstrap bounds imply that at least for some range of $\Df$ the lowest dimension operator in a consistent CFT correlator must be a scalar.
It follows then that:
\bea
\mathcal (z \bar z)^{\Df}\mathcal G(\rho,\bar \rho)\sim 1+ a_{\Delta_g,0}\,R^{\Delta_g} +\ldots\label{eq:approx}
\eea
At this point a clarification is in order: we are after all aiming to place bounds on the space of all consistent correlators, and an arbitrary generic correlator could have very closely space operators, or even a continum.  Who is to say we should take the above to be a good approximation? The reason is that what we are really interested in characterizing here are the properties of the maximal value or minimal value correlators, not generic ones. Since such correlators saturate bounds, they come along with associated linear functionals (i.e. they are extremal  \cite{El-Showk:2012vjm,El-Showk:2016mxr}) and hence they must have a sparse spectrum. In particular such correlators are indeed well approximated by a relation such as the one above for sufficiently small $R$.

With this very simple approximation, we begin by addressing the question of an upper bound on the valueof a correlator. The approximation \reef{eq:approx} teaches us several things. Firstly, since $R$ is small in the fundamental domain, we can see that in order to maximize the value of the correlator we would do well to lower the gap as much as possible. Hence, in the set of correlators whose non-identity scalar operators all have dimensions above some $\Delta_g$ we expect that the maximal value correlator should necessarily have a scalar operator at, or very close to, the gap $\Delta_g$. We also expect that in general lowering the gap leads to higher maximum values. Secondly, to maximize the correlator we would like to increase the OPE of that low-lying operator, $a_{\Delta_g,0}$ as much as possible. Hence we expect that:
\bea
\GM(z,\bar z|\Delta_g)\gtrapprox \mathcal G_{\mbox{\tiny opemax}}(z,\bar z|\Delta_g) \label{eq:expectsup}
\eea
That is, the problems of correlator maximization and OPE maximization are related, at least for small enough values of $R$.

Now let us see what we can say about lower bounds on the correlator, turning again to the approximation \reef{eq:approx} for guidance. In this case the logic is reversed, so that to minimize the value of the correlator it seems that we should increase $\Delta_g$ and minimize $a_{\Delta_g,0}$ as much as possible. Roughly speaking, we expect that OPE coefficients should decrease as the dimension increases, so it seems like both these effects push up the gap as much as possible. An important point is that what this argument tells us is that we expect that minimizing correlator should have a large gap {\em across all spin sectors}. It just so happens that there is an upper bound on the scalar gap, and for a range of $\Df$ this upper bound $\Delta_{\mbox{\tiny gapmax}}$ is smaller than the unitarity bound on the spin-2 sector $\Delta_u(2)=d$. Under these conditions, we expect
\bea
\Gm(w,\bar w)\lessapprox \mathcal G_{\mbox{\tiny gapmax}}(w,\bar w)\,, \qquad \mbox{as long as}\quad \Delta_{\mbox{\tiny gapmax}}<d\,. \label{eq:expectinf}
\eea
So again we find a relation between the two a priori distinct problems of correlator minimization and gap maximization.

Although strictly speaking both these expectations should be borne out only for very small $R$, in practice we will see they are often true even for $R=O(1)$.

\subsection{Exact results for $1d$ kinematics}
\label{sec:exact}
In recent work \cite{Paulos:2020zxx}, one of us has proven exact bounds on CFT correlators restricted to the line $z=\bar z$. Let us set $\mathcal G(w)\equiv \mathcal G(w,\bar w)$ the line restriction of an arbitrary CFT correlator in spacetime dimension $d\geq 1$. We can write it as:
\bea
\mathcal G(w)=\sum_{\Delta\geq 0} a_{\Delta}^{1d} G_{\Delta}(w|\Df),\qquad G_{\Delta}(w|\Df)=w^{\Delta-2\Df}\ _2F_1(\Delta,\Delta,2\Delta,w)\,.
\eea
We have expanded the correlator in terms of SL(2,$\mathbb R)$ primary operators, with $G_{\Delta}$ an $SL(2,\mathbb R)$ conformal block.  For correlators arising from a CFT$_d$, this $SL(2,\mathbb R)$ spectrum contains both higher-$d$ primaries as well as higher-$d$ descendant states. Accordingly there will be relations among the effective $SL(2,\mathbb R)$ OPE coefficients, which we have denoted $a_{\Delta}^{1d}$. The Polyakov bootstrap, proposed in \cite{Sen:2015doa,Gopakumar:2016wkt,Gopakumar:2016cpb,Gopakumar:2018xqi} and proven for $d=1$ CFTs in \cite{Mazac:2018ycv} tells us we may write
\bea
\mathcal G(w)=\sum_{\Delta} a_{\Delta}^{1d}\, \mathcal P^B_{\Delta}(w)
\eea
or alternatively
\bea
\mathcal G(w)=\sum_{\Delta}a_{\Delta}^{1d}\, \mathcal P^F_{\Delta}(w).
\eea
Here $\mathcal P^B_{\Delta}$, $\mathcal P^F_{\Delta}$ are called the bosonic and fermionic Polyakov blocks. They are crossing symmetric functions and can be obtained as certain sums of Witten exchange diagrams and contact terms for bosonic and fermionic theories in AdS$_2$.\footnote{It is important however that the above are rigorous identities that must hold for any unitary CFT, and in this sense have nothing to do with holography or perturbation theory.} Alternatively, the Polyakov blocks are computable from the action of certain linear functionals $\Omega^{B,F}_w$, called master functionals in \cite{Paulos:2020zxx}:
\bea
\pm \mathcal P_{\Delta}^{B,F}(w)=\Omega_w^{B,F}[F_{\Delta}]-G_{\Delta}(w|\Df)
\eea
where $F_{\Delta}(w)\equiv G_{\Delta}(w|\Df)-G_{\Delta}(1-w|\Df)$. This representation of the Polyakov blocks leads to the following expressions:
\bea
\mathcal P^{B}_\Delta(w)&=-\sin^2\left[\frac{\pi}2(\Delta-2\Df)\right]\int_0^1 \ud z g^B_w(z) G_{\Delta}(z|\Df)\,,&\qquad \Delta&> 2\Df \\
\mathcal P^{F}_\Delta(w)&=+\cos^2\left[\frac{\pi}2(\Delta-2\Df)\right]\int_0^1 \ud z g^F_w(z) G_{\Delta}(z|\Df)\,,& \qquad \Delta&> 2\Df-1
 \eea
where crucially, the master functional kernels $g^F_w(z)$ and $g^B_w(z)$ are positive for $w,z\in(0,1)$. Since the SL(2,$\mathbb R)$ blocks are also positive in that range, the Polyakov bootstrap together with the above establishes the bounds:

\bea
\overline{\mathcal G}(w)&:=\mathcal G(w)-\sum_{0\leq \Delta\leq 2\Df} a_{\Delta}^{1d}\, \mathcal P^B_{\Delta}(w)\leq 0\,,\\
\underline{\mathcal G}(w)&:=\mathcal G(w)-\sum_{0\leq \Delta\leq 2\Df-1} \!\!\!\!a_{\Delta}^{1d}\, \mathcal P^F_{\Delta}(w)\geq 0
\eea
valid for $w\in (0,1)$. In particular the identity Polyakov blocks, which always appear in the subtractions above, are computable as:
\ba
\mathcal P^B_{\Delta}(w)=\mathcal G^B(w)\equiv \frac 1{w^{2\Df}}+\frac 1{(1-w)^{2\Df}}+1\\
\mathcal P^F_{\Delta}(w)=\mathcal G^F(w)\equiv \frac 1{w^{2\Df}}+\frac 1{(1-w)^{2\Df}}-1
\ea
which are bosonic / fermionic generalized free field correlators, respectively. The fermionic correlator makes sense only for $d=1$ as it cannot be extended for $z\neq \bar z$. However the bosonic can:
\ba
\mathcal G^B(w)=\mathcal G^B(w,\bar w)\bigg|_{w=\bar w}\,, \qquad \mathcal G^B(w,\bar w)=\frac 1{(w \bar w)^\Df}+\frac 1{[(1-w)(1- \bar w)]^\Df}+1\,.
\ea
Finally, although it is not guaranteed by the above, by an examination of the master functional action it was observed that
\ba
\mathcal P_{\Delta}^B(w)&\geq 0\,,& \qquad 0\leq \Delta&\leq 2\Df\\ \mathcal P_{\Delta}^F(w)&\geq 0\,,& \qquad 0\leq \Delta&\leq  2\Df-1
\ea
Putting everything together we get:
\ba
\mathcal G(w)\equiv \mathcal G(w,w)&\geq \mathcal G^F(w)\\
\mathcal G(w)\equiv \mathcal G(w,w)&\leq \mathcal G^B(w)\,, \qquad \mbox{if}\quad \Delta_g\geq 2\Df\,.
\ea
As a couple of comments, notice that interestingly for $d=1$ the  expectations \reef{eq:expectinf} \reef{eq:expectsup} quoted above are actually exact statements for any $w$, i.e.:
\ba
\Gm^{\mbox{\tiny (1d)}}(w)&=\mathcal G_{\mbox{\tiny gapmax}}^{\mbox{\tiny (1d)}}(w)=\mathcal G^F(w)\,,\qquad \GM^{\mbox{\tiny (1d)}}(w|2\Df)&=\mathcal G^{\mbox{\tiny (1d)}}_{\mbox{\tiny opemax}}(w|2\Df)=\mathcal G^B(w)
\ea
The second comment is that since the generalized free boson is a good solution to crossing in arbitrary spacetime dimension, the upper bound quoted above is actually optimal. Although it is restricted to the line, we will give numerical evidence that it holds throughout the Euclidean section. Analytically though, we can only obtain a weaker result, as we now demonstrate.

The positivity properties of  expansion \reef{eq:blockexp} imply
\bea
G_{\Delta,\ell}(\rho,\bar \rho)\leq G_{\Delta,\ell}(R,R).
\eea
Let us set
\bea
\mathcal G_\rho(R,R)=\mathcal G_z(z_{\mbox{\tiny eff}},z_{\mbox{\tiny eff}})\,, \qquad \mbox{with}\quad z_{\mbox{\tiny eff}}\equiv \frac{4R}{(1+R)^2}
\eea
This implies:
\begin{multline}
\mathcal G(\rho,\bar \rho)=\frac{\sum_{\Delta,\ell} a_{\Delta,\ell} G_{\Delta,\ell}(\rho,\bar \rho)}{(z\bar z)^{\Df}} \leq \frac{\sum_{\Delta,\ell} a_{\Delta,\ell} G_{\Delta,\ell}(R,R)}{(z\bar z)^{\Df}} \\
=\left[1-z_{\mbox{\tiny eff}}\,\sin^2\left(\mbox{$\frac{\phi}2$}\right)\right]^{2\Df} \mathcal G(R,R)\,.
\end{multline}
This is better than the coarser $\mathcal G(\rho,\bar \rho)\leq \mathcal G(R,R)$ which follows from the Cauchy-Schwarz inequality for the correlator. Altogether this implies
\bea
\mathcal G(\rho,\bar \rho)\leq \left[1-z_{\mbox{\tiny eff}} \sin^2\left(\mbox{$\frac{\phi}2$}\right)\right]^{2\Df} \left(1+\frac{1}{z_{\mbox{\tiny eff}}^{2\Df}}+\frac{1}{(1-z_{\mbox{\tiny eff}})^{2\Df}}\right)\,,  \qquad \mbox{if}\quad \Delta^{\mbox{\tiny gap}}_{\mathcal G}\geq 2\Df
\eea
with the bound being optimal for $\phi=0$. A caveat is that the bound holds only if there is a gap on the full set of SL(2,$\mathbb R$) primaries. This means that for higher-$d$ CFT correlators restricted to the line the gap must be imposed uniformly across all spins. In particular the bound is less interesting when $\Df>d/2$ as it would rule out CFT correlators with a stress-tensor.

\subsection{Lorentzian kinematics}

We now discuss the case where we have spacelike separated operators but we are not in the Euclidean section, so that $z,\bar z$ are both real and independent variables in the fundamental domain where $z,\bar z\in (0,1)\times (0,1)$. This regime corresponds to taking imaginary angle in the polar representations of section \ref{sec:intuition}, so that e.g. $\rho=R e^{-\phi}$ with real $\phi$. In particular the argument of the Gegenbauer polynomials in \reef{eq:blockexp} becomes $\cosh \phi$. This means then that our basic approximation \reef{eq:approx} is in general not valid. This is because although the radius $R$ remains small inside the fundamental domain, the Gegenbauer polynomials are now exponentially growing in spin and this may compensate the exponential suppression with respect to $R$. The only way to guarantee this does not occur is to keep both $\rho$ and $\bar \rho$ small and $\rho\sim \bar \rho$. In this regime our expectations for correlator minimization and maximization are the same as our discussion in the Euclidean regime.

The more interesting case corresponds to taking one of the cross-ratios to be relatively large, i.e. approaching one. Notice that inside the fundamental domain this necessarily means that the other must be small, since e.g. $\sqrt{\rho \bar \rho}\leq 3-2\sqrt{2}$ inside this region. In this case it is better to use a different series expansion for the conformal block:
\bea
G_{\Delta,\ell}(z,\bar z)=z^{\Delta-\ell} \bar z^{\Delta-\ell}\sum_{n,m=0}^{\infty} c_{n,m} z^{n} \bar z^m\,,\qquad c_{n,m}\geq 0\,, \label{eq:blockexplor}
\eea
where $c_{n,m}\geq 0$ follows from unitarity \cite{Pappadopulo:2012jk}. In particular, for small $z$ we have
\bea
G_{\Delta,\ell}(z,\bar z)\sim c_{0,2\ell} z^{\Delta-\ell} \bar z^{\Delta+\ell}\,_2F_1\left(\frac{\Delta+\ell}2,\frac{\Delta+\ell}2,\Delta+\ell,\bar z\right)
\eea
so that in this regime, $z\ll \bar z \sim 1$, the parameter controlling the operators which give dominant contributions to the correlator is the twist $\Delta-\ell$.

Let us suppose firstly that $d\geq 3$. In this case unitarity allows in principle for scalar operators to dominate since the lowest allowed twists are $\frac{d-2}2$ and $d-2$ for spin zero and greater than zero respectively. In particular, if we consider the correlator maximization problem with a scalar gap $\Delta_g$ which is not too close to $d-2$, we expect that the result should be essentially the same as in the Euclidean case: the maximal value correlator should essentially be the same as the opemax correlator for an operator which sits at the gap. If in contrast $d=2$, or if we choose to set the gap above $d-2$ then in principle the only thing we can say is that correlator maximization should involve minimizing the twist gap across all spins, and secondly maximizing their overall contribution, by increasing their OPE coefficients, but unfortunately we cannot make any general statement. 

The same logic applies to correlator minimization. In this case the gapmax correlator always has a scalar operator whose twist is above $d-2$ (for instance when $\Df=\frac{d-2}2$ the gap sits at $d-2$ and is saturated by free theory). Hence for this problem we always expect that the solution will involve increasing the overall twist gap across all spins but not necessarily uniformly across spins. 

After these general remarks, we will now establish rigorous bounds on the correlator in the Lorentzian regime in terms of those in $d=1$. This can be done quite trivially given expansion \reef{eq:blockexplor} since it implies
\bea
G_{\Delta,\ell}(z_m,z_m)\leq G_{\Delta,\ell}(z,\bar z) \leq G_{\Delta,\ell}(z_M,z_M), \qquad z_m=\mbox{min}\{z,\bar z\},\quad z_M=\mbox{max}\{z,\bar z\}
\eea
and therefore
\bea
z_m^{2\Df}\mathcal G(z_m,z_m) \leq (z\bar z)^\Df\mathcal G(z,\bar z)\leq  z_M^{2\Df}\mathcal G(z_M,z_M)\,.
\eea
Using the $d=1$ bounds we find for $z,\bar z\in (0,1)$:
\ba
 \mathcal G(z,\bar z)&\geq \left(\frac{z_m}{z_M}\right)^{\Df} \Gm^{(d)}(z_m,z_m)\geq \left(\frac{z_m}{z_M}\right)^{\Df}\mathcal G^{F}(z_m)\,\\
\mathcal G(z,\bar z)&\leq \left(\frac{z_M}{z_m}\right)^{\Df}\GM^{(d)}(z_M,z_M|\Delta_g)\leq \left(\frac{z_M}{z_m}\right)^{\Df}
\GM^{(d=1)}(z_M|\Delta_g)\qquad \Delta_g^{\mathcal G}\geq \Delta_g
\ea
Using these we can study the double lightcone limit, where $z\to 0 $ and $\bar z\to 1$ at the same rate. In this case we write
$
z_m=z, z_M=1-z
$
and take the limit where $z\to 0$. Then we can find
\ba
z^{\Df}\mathcal G(z,1-z)&\underset{z\to 0^+}\geq 1 \\
z^{3\Df}\mathcal G(z,1-z)&\underset{z\to 0^+}{\leq} 1\,
\ea
The lower bound cannot be parametrically improved, since it is satisfied by generalized free fields. As for the upper bound, notice that it holds independently of the choice of gap (as long as the maximal value correlator exists, i.e. it is not infinite). We believe however that in this case it can be parametrically improved to $\mathcal G(z,1-z)=O(z^{-2\Df})$, which holds for $d=2$ CFTs \cite{Collier:2018exn,Paulos:2020zxx}.

\section{Correlator bounds from linear optimization}
\label{sec:opt}
\subsection{Formulation of the linear programming problems}
Consider conformal correlators admitting an expansion of the following form:
\bea
\mathcal G(w,\bar w)=\frac{1}{(w\bar w)^\Df}+\sum_{\Delta,\ell\in \mathcal S} a_{\Delta,\ell} G_{\Delta,\ell}(w,\bar w|\Df)
\eea
with corresponding crossing equation
\ba
F_{0,0}(w,\bar w)+\sum_{\Delta,\ell\in \mathcal S} a_{\Delta,\ell} F_{\Delta,\ell}(w,\bar w|\Df)=0\,. \label{eq:cross2}
\ea
The first term stands for the identity contribution which we have singled out. The set $\mathcal S$ should be some subset of $\mathcal U$, the set of those quantum numbers $\Delta,\ell$ allowed by unitarity, namely:
\ba
\mathcal U=\left\{(\Delta,\ell):  \left.\begin{array}{ll}
	\ell=0:& \Delta\geq \frac{d-2}2\\
	\ell\in 2 \mathbb N: & \Delta\geq d-2+\ell
	\end{array}
\right.
\right\}
\ea
Often we will simply set $\mathcal S=\mathcal U$, othertimes we will impose additional gaps in $\Delta$.

We will consider both minimization and maximization problems so as to get lower and upper bounds on values of correlators on the Euclidean section. Such bounds can be obtained by constructing linear functionals that act on the crossing equation \reef{eq:cross2}, satisfying suitable properties. We will denote functionals providing upper/lower bounds at a given point $w,\bar w$ as $\Lu$ and $\Ll$. The action of such a functional on a crossing vector $F_{\Delta,\ell}$ is denoted for instance $\Lu(\Delta,\ell)$. The conditions to be imposed are simply
\bea
\underline \Lambda_{w,\bar w}(\Delta,\ell)\geq -G_{\Delta,\ell}(w,\bar w|\Df)\,\qquad \overline \Lambda_{w,\bar w}\geq G_{\Delta,\ell}(w,\bar w|\Df)\,\,,\qquad \mbox{for all}\quad \Delta,\ell \in \mathcal S\,.
\eea
Acting with these functionals on the crossing equation we find:
\ba
\Ll(0,0)&=-\sum_{\Delta,\ell\in \mathcal S} a_{\Delta,\ell}\Ll(\Delta,\ell)\leq \mathcal G(w,\bar w)-\frac{1}{(w\bar w)^\Df}\\
-\Lu(0,0)&=+\sum_{\Delta,\ell\in \mathcal S} a_{\Delta,\ell}\Lu(\Delta,\ell)\geq \mathcal G(w,\bar w)-\frac{1}{(w\bar w)^\Df}\,.
\ea
This implies:
\bea
\frac{1}{(w\bar w)^\Df}+\Ll(0,0)\leq\mathcal G(w,\bar w)\leq \frac{1}{(w\bar w)^\Df}-\Lu(0,0)\,.
\eea
We are now ready to formulate our optimization problems:\\
\vspace{0.3 cm}\\
{\bf Correlator minimization:}
\ba
\underset {\Ll\in \mathcal L}{\mbox{max}} \ \Ll(0,0) \quad \mbox{s.t.}\qquad
\Ll(\Delta,\ell)\geq -G_{\Delta,\ell}(w,\bar w|\Df)\quad \mbox{for all}\ (\Delta,\ell)\in \mathcal S
\ea
\vspace{0.3 cm}\\
{\bf Correlator maximization:}
\ba
 \underset{\Lu\in \mathcal L}{\mbox{max}} \ \Lu(0,0) \quad \mbox{s.t.}\qquad
\Lu(\Delta,\ell)\geq G_{\Delta,\ell}(w,\bar w|\Df)\quad \mbox{for all}\ (\Delta,\ell)\in \mathcal S
\ea
In the above the set $\mathcal L$ is typically some finite dimensional but large search space of elementary basis functionals. The idea therefore is to look for functionals satisfying the right positivity conditions inside $\mathcal L$, choosing the one whose action on the identity is the largest. In section \ref{sec:numapp} below we use the tried and tested approach of choosing $\mathcal L$ to be a finite set of derivatives at the crossing-symmetric point $z=\bar z=\frac 12$, leaving an exploration of better functional bases \cite{Mazac:2018ycv,Mazac:2018mdx,Paulos:2019gtx,Paulos:2019fkw,Mazac:2019shk} for the future.

Let us obtain some basic bounds following this logic. Consider first minimization. A simple functional satisfying all the desired conditions is simply
\bea
\Ll[F_{\Delta,\ell}]=-F_{\Delta,\ell}(w,\bar w)=G_{\Delta,\ell}(1-w,1-\bar w)-G_{\Delta,\ell}(w,w|\Df)
\eea
which leads to the (trivial) bound
\ba
\mathcal G(w,\bar w)\geq \mbox{max}\left\{\frac{1}{(w \bar w)^{\Df}}, \frac{1}{[(1-w)(1-\bar w)]^{\Df}},1\right\}
\ea
where in passing we have used crossing symmetry to get better bounds for all $w,\bar w$ where the correlator is real-valued. This bound follows simply  from the statement that the correlator is the sum of the identity block in some OPE channel plus positive contributions.

To get an upper bound we have to work slightly harder. Let us obtain a bound at the crossing symmetric point $w=\bar w=\frac 12$, choosing:
\bea
\overline{\Lambda}_{\frac 12,\frac 12}[F_{\Delta,\ell}]=\frac \lambda 2 \partial_{z} F_{\Delta,\ell}(z,z)\bigg|_{z=\frac 12}=\lambda \partial_{z} G_{\Delta,\ell}(z,z)\bigg|_{z=\frac 12}
\eea
It can be checked that in general this will only satisfy the required constraints above some gap depending on $\lambda$. This is in line with our expectations that we do not expect a bound to exist for general $\Df$ unless we set a gap in the spectrum. To get a nice result let us work in the limit of large $\Df$, where conformal blocks at large $\Delta$ take the form
\bea
G_{\Delta,\ell}(w,\bar w|\Df)\underset{\Delta,\Df\to \infty}\sim \frac{(4\rho)^{\Delta}}{z^{2\Df}} g_\ell(w,\bar w)
\eea
where $g_{\ell}(w,\bar w)$ does not vary exponentially with $\Delta$. Let us demand that our functional provides a bound on correlators for which the gap satisfies $\Delta_g\geq \alpha \Df$. The optimal such functional will satisfy
\bea
\overline{\Lambda}_{\frac 12,\frac 12}(\Delta_g,\ell)=G_{\Delta_g,\ell}(\mbox{$\frac 12,\frac 12$}|\Df)
\eea
i.e. the positivity constraints will be saturated at the gap value. Note that the value of $\ell$ is irrelevant in the above, since $g_{\ell}$ is independent of $\ell$ when $w=\bar w$ in the large $\Delta$ limit. We can use this to fix $\lambda$:
\bea
\lambda=\frac{1}{2\sqrt{2} \Df} \frac{1}{\alpha-\sqrt{2}}
\eea
One can check that the functional will satisfy the right positivity conditions above the gap, but only if $\alpha>\sqrt{2}$. From this we get the bound
\bea
\mathcal G(\mbox{$\frac 12,\frac 12$})\underset{\Df\to \infty}{\leq} 2^{2\Df}\left(1+\frac{\sqrt{2}}{\alpha-\sqrt{2}}\right)\,, \qquad \Delta_g^{\tiny \mathcal G}\geq \alpha \Df>\sqrt{2}\Df
\eea
As an example, setting $\alpha=2$ we know the optimal bound correspond to the generalized free boson solution, which scales as $2^{2\Df}\times 2$. The above is parametrically as good as the optimal bound, scaling $2^{2\Df}\times (2+\sqrt{2})$. This result does not give us any bound if the gap is smaller than $\sqrt{2}\Df$. Could this be improved? We believe the answer is no. Firstly, it can be checked that including an arbitrary number of derivatives (but smaller than $\Df$) won't change this conclusion. Secondly, the analytic functionals and corresponding bounds on OPE coefficients determined in \cite{Mazac:2018mdx} also precisely break down at that point. So it is rather nice that such a simple functional already correctly captures this feature.

\subsection{Aside: reformulating the gap maximization problem}

Let us make contact with our expectation that the correlator minimization problem is  related to gap maximization. A similar argument to the one below can be made for the relation between correlator maximization and OPE maximization. 

Suppose one wanted to prove a bound on the leading scaling dimension in the spin channel $\ell=\ell_0$. One can do this by constructing a functional, $\beta$, inside some finite dimensional search space $\mathcal L$, satisfying the constraints:
\ba
\beta(0,0)&=0\,,& \qquad &\partial_{\Delta} \beta(\Delta_g,\ell_0)=1\,,\\
\beta(\Delta,\ell_0)&\geq 0&\qquad& \mbox{for} \quad \Delta> \Delta_g\\
\beta(\Delta,\ell)&\geq 0&\qquad &\mbox{for} \quad (\Delta,\ell) \in \mathcal S\,, \quad \ell\neq \ell_0
\ea
If such a functional can be found, it is easily seen (by acting with $\beta$ on the crossing equation) that it establishes an upper bound on the dimension equal to $\Delta_g$. For each choice of $\Delta_g$, finding whether the corresponding $\beta$ exists is an optimization problem. In practice, one must fix $\Delta_g$, check if such a functional exists, lather, rinse, repeat, until one finds the smallest such $\Delta_g$ possible. We denote the lowest such $\Delta_g$ by $\Delta_{\mbox{\tiny gapmax}}$: it is the best possible bound available on the gap, given the constraints in set $\mathcal L$.

Fortunately, there is a better way. Consider the correlator minimization functional. At optimality we have
\ba
\frac{1}{(w\bar w)^\Df}+\Ll(0,0)=\mathcal G_{\mbox{\tiny min};w,\bar w}(w,\bar w)\\
\ea
Let us assume that the lowest dimension, non-identity operator in $\mathcal G_{\mbox{\tiny min};w,\bar w}(w,\bar w)$ is a scalar of dimension $\Delta_g^{r}$, with $r=\sqrt{w\bar w}$. In the limit where $r\to 0$ we get
\ba
\Ll(0,0)&\underset{r\to 0}{\sim} a_{\Delta_g^{r},0}\, r^{\Delta_g^{r}-2\Df}\\
\Ll(\Delta,\ell)&\underset{r\to 0}{\geq} -r^{\Delta-2\Df} C^{\left(\frac{d-2}2\right)}_{\ell}(\cos \theta)
\ea
with the inequality being saturated at $\Delta=\Delta_g^r$.
It follows that the functional 
\be
\beta\equiv \lim_{r\to 0}\frac{\Ll}{\partial_{\Delta} \Lambda_{w,\bar w}(\Delta_g^r,0)}
\ee
satisfies
\ba
\beta(0,0)=0\,,\quad \partial_\Delta \beta(\lim_{r \to 0}\Delta_g^r,0)=1\,, \quad  \beta(\Delta,\ell)\geq 0\,,\quad \mbox{for}\quad \Delta> \lim_{r\to 0} \Delta_g^{r}
\ea
Hence $\beta$ is a valid gap maximization functional, and leads to:
\bea
\Delta_{\mbox{\tiny gapmax}}=\lim_{r\to 0} \Delta_g^{r}
\eea
The conclusion is that minimizing the correlator for $r\to 0$ is the same as gap maximization as long as the leading operator is a scalar. We will see direct evidence for this in the numeric applications of section \ref{sec:numapp}.

This result also suggests a new way to maximize the gap by a single optimization step. It is not really necessary to minimize the full correlator, since the important point really is to introduce constraints that decay sufficiently fast with $\Delta$. It is also possible to do away with the restriction that the correlator minimization agrees with gap maximization only when the leading operator is a scalar, and to be able to maximize the gap in other spin sectors. To do this consider then the following optimization problem:
\ba
\underset{\omega\in \mathcal L}{\mbox{max}} \ \omega(0,0) \quad \mbox{s.t.}\qquad \left.\begin{array}{l}
		\omega(\Delta,\ell_0)\geq -c(\Delta)\,,\qquad  \Delta>\Delta_0\\
		\\
\omega(\Delta,\ell)\geq 0 \qquad \mbox{for}\quad (\Delta,\ell)\in U, \quad \ell\neq \ell_0
\end{array}
\right.\,.
\ea
Here $c(\Delta)$ should be a non-negative function which should decay quickly with $\Delta$ and be positive at least in some range where the maximal gap is expected to occur. Good choices are for instance:
\ba
\mbox{Exponential:}&\qquad& c(\Delta)&=\epsilon^\Delta\,, \qquad \epsilon\ll 1\\
\mbox{Power law:}&\qquad& c(\Delta)&= \frac{1}{(\Delta-\Delta_0)^{p}}
\ea
The optimal value for this optimization problem will be given by a functional $\omega^{\mbox{\tiny opt}}$ dual to an approximate solution to crossing. This solution minimizes the cost in the following dual problem:
\ba
 \underset{a_{\Delta,\ell}}{\mbox{min}} \sum_{\Delta>\Delta_0} a_{\Delta,\ell_0} c(\Delta)\quad \mbox{s.t.}\qquad \sum_{\substack{\ell\neq \ell_0: (\Delta,\ell)\in \mathcal U\\\ell=\ell_0: \Delta>\Delta_0}} a_{\Delta,\ell} \,\omega(\Delta,\ell)=0 \quad \mbox{for all}\quad \omega\in\mathcal L
\ea
From this latter perspective it is clear  that the outcome of the problem is a solution with as large a gap as possible while remaining consistent with the functional constraints. Denoting  $\Delta_g^{\mbox{\tiny min}}$ the leading scaling dimension (in channel $\ell_0$) of the solution to this problem, it is clear that
\bea
\Delta_g^{\mbox{\tiny min}}\sim \Delta_{\mbox{\tiny gapmax}}
\eea
Depending on our choice $c(\Delta)$ we can get an arbitrarily good estimate for $\Delta_{\mbox{\tiny gapmax}}$. This is achieved by setting $\epsilon\ll 1$ in the exponential ansatz, or $p\gg 1$ and $\Delta_0$ as close as possible to the expected maximal gap. Once we have such an estimate, we can then simply check that a functional $\beta$ exists for $\Delta_g=\Delta_g^{\mbox{\tiny min}}+\eta$, for small $\eta$. This can be done very efficiently 
since we can ``hotstart'' the search for the $\beta$ functional with $\omega^{\mbox{\tiny opt}}$. Alternatively, we can simply ``flow'' from the above problem to the gap maximization problem \cite{El-Showk:2016mxr}: essentially, the $\beta$ functional can be immediately constructed merely from knowledge of the solution to the dual problem, without any new optimization required to be solved. 

These approaches are all equivalent, and all lead to an extremal functional $\beta$ with the correct properties establishing indeed that $\Delta_g^{\mbox{\tiny min}}=\Delta_{\mbox{\tiny gapmax}}$ to any desired accuracy. We have tested this procedure using both an exponential cost function as well as a power law of high degree. For instance, setting $\epsilon=10^{-150}$ we obtained a $\Delta_{\mbox{\tiny gapmax}}-\Delta_g^{\mbox{\tiny min}}<10^{-15}$, which we checked by explicitly constructing a gapmax functional after obtaining the initial estimate. We believe this method will be useful for future numerical bootstrap applications.

\section{Numerical applications}
\label{sec:numapp}

In this section we present numerical bounds on 3d CFT correlators. The bounds are obtained by applying the logic of the previous section, constructing suitable functionals inside some search space using the  {\tt JuliBootS} package \cite{Paulos:2014vya}. Further details on the numerical implementation are relegated to appendix~\ref{detail}.

\subsection{Summary}

\begin{figure}[ht]
  \centering
	\includegraphics[width=15cm]{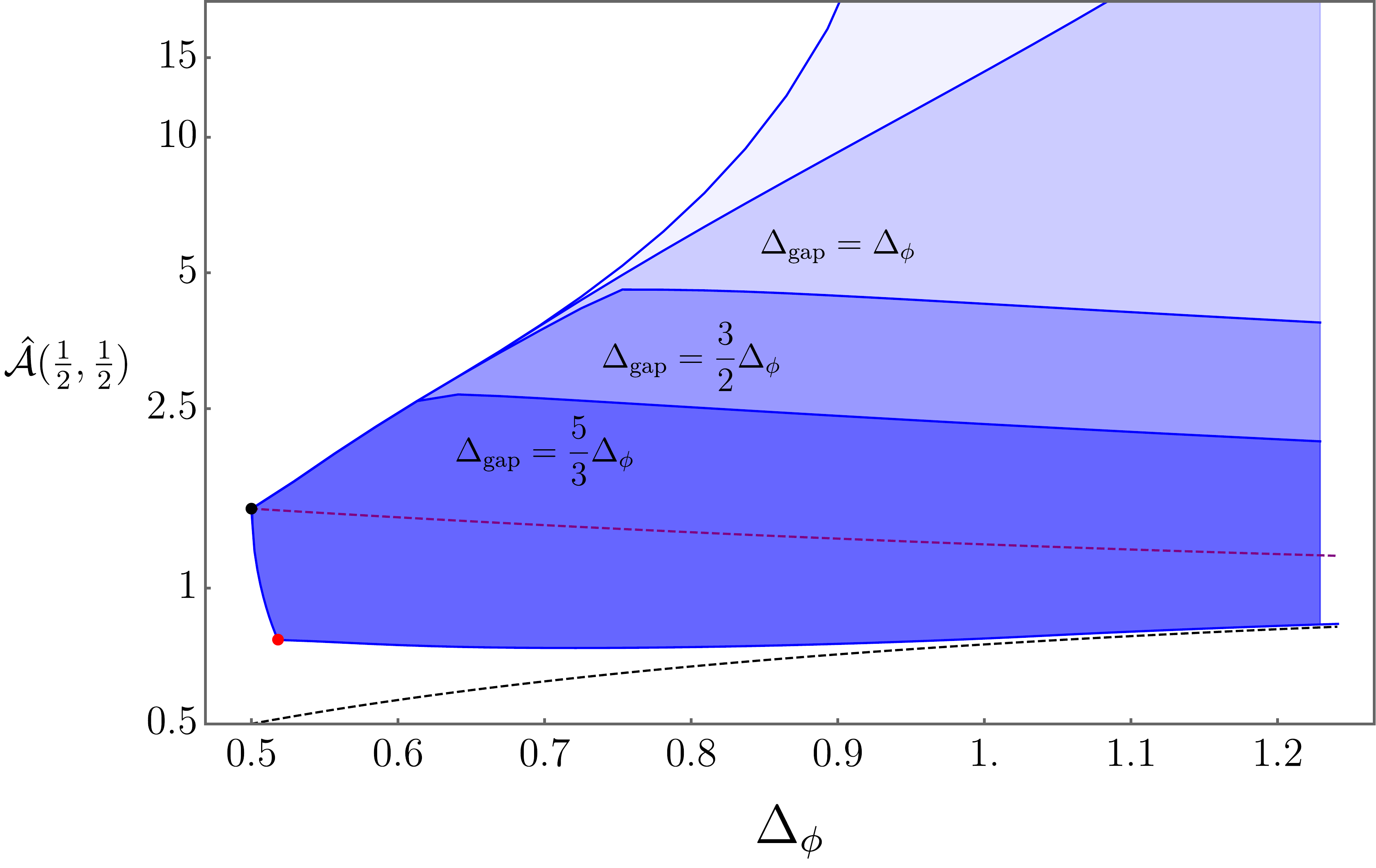}
\caption{\label{allowed}
	Bounds on 3d CFT correlator values, with $\hat{\mathcal A}(z,\bar z)=(z\bar z)^{2\Df}\mathcal G(z,\bar z)-1$. The shaded region represents the values such a correlator may take at the crossing symmetric point $z=\bar z=\frac 12$. The dashed lines inside the allowed region correspond to maximal allowed values assuming a gap. The upper bound goes to $\infty$ close to $\Df=1$. The dashed line outside the allowed region is the 1d generalized free fermion correlator, which provides a (suboptimal) lower bound. The dashed line inside the allowed region is the generalized free boson correlator, which provides an upper bound for $\Delta_{\mbox{\tiny gap}}=2\Df$. The black and red dots are the free theory and 3d Ising values respectively.
}
\end{figure}

\begin{figure}[ht]
\centering
	\includegraphics[width=15cm]{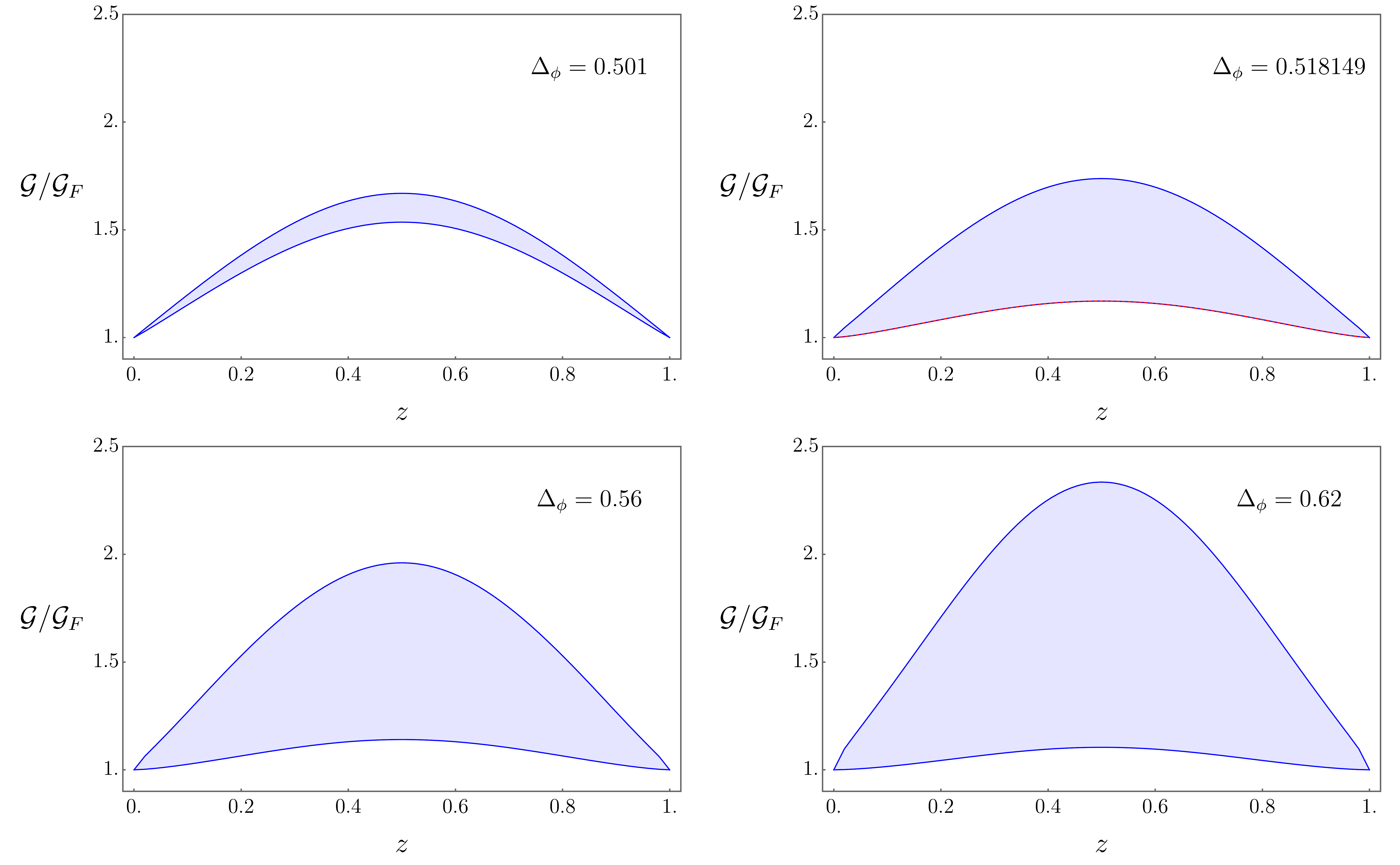}
\caption{\label{allowed1}Allowed values for CFT correlators on the line $z=\bar z$, for various values of $\Df$, normalized by the 1d generalized free fermion correlator. The 3d Ising correlator matches the lower bound for $\Df=\Delta_{\sigma}^{\mbox{\tiny Ising}}\sim 0.518149$.}
\end{figure}

Our numerical results are summarized in figure \ref{allowed},  which shows the allowed range of values of the four-point correlator at the crossing symmetric point $z=\bar z=\frac 12$ as a function of $\Df$, and figure \ref{allowed1}, which shows upper and lower bounds along the line $z=\bar z$ for various values of $\Df$. There are several interesting features in these plots, so let us discuss them in turn. 

Beginning from the top of figure~\ref{allowed}, we see that there is an upper bound on the correlator, at least in some range of $\Df$. This bound is valid without any particular assumptions on the spectrum. In the figure we also show the effect of imposing various gaps in the scalar sector, leading to stronger bound curves. The upper bound assuming a gap of $2\Df$ is closely saturated by the exact generalized free boson correlator, in agreement with the exact bound reviewed in section~\ref{sec:exact}. In the absence of an assumption in the gap, the upper bound diverges in the vicinity of $\Df=1$. This can be understood from the fact that there is a unitary solution to the crossing equation without identity for $\Df\geq d-2$, namely
\bea
\mathcal G_C(u,v)=(u v)^{-\frac{\Df}2}+u^{-\frac{\Df}2}+v^{-\frac{\Df}2}
\eea
This correlator arises in the computation of the $\phi^2$ 4-point function in the generalized free CFT. One can check by direct expansion that unitarity of this solution is violated unless $\Df\geq d-2$. Since this solution may be added to any given correlator with an arbitrarily large coefficient, there can be no bound without assuming a gap in the spectrum for such values of $\Df$. In particular the gap $\Delta_g$ must be necessarily larger than $\Df/2$. 

As we move towards the left along the top curve we hit a kink which occurs at the free theory point, where $\Df=1/2$. The upper and lower bounds reassuringly coincide both with each other and the free theory value, as the only CFT correlator with that dimension should be the free one. Continuing now below and to the right, we have a lower bound on the correlator. It is compatible with the exact lower bound on the correlator determined by the 1d generalized free fermion solution reviewed in section~\ref{sec:exact}. Although our bound is stronger since it takes into account constraints from the full 3d crossing equation, nevertheless, it seems to rapidly approach the 1d bound for moderate values of $\Df$. 

The lower bound has a distinctive kink at $\Df\sim 0.518149$, which is the dimension of the spin field in the 3d Ising CFT. This is not an accident. Indeed, in figure \ref{allowed1} we see that at that particular value of $\Df$ the infimum correlator actually matches the 3d Ising correlator \cite{Simmons-Duffin:2016wlq} extremely accurately. This is in perfect agreement with our argument that correlator minimization is closely related to gap maximization, since gap maximization indeed leads to the 3d Ising model at the correct value of $\Df$ \cite{El-Showk:2012cjh}. We will further confirm this below by a comparison of the leading scalar dimension in gap maximization versus correlator minimization.

\subsection{Minimization}
\label{sec:min}

\begin{figure}[ht]
	\centering
	\includegraphics[width=15cm]{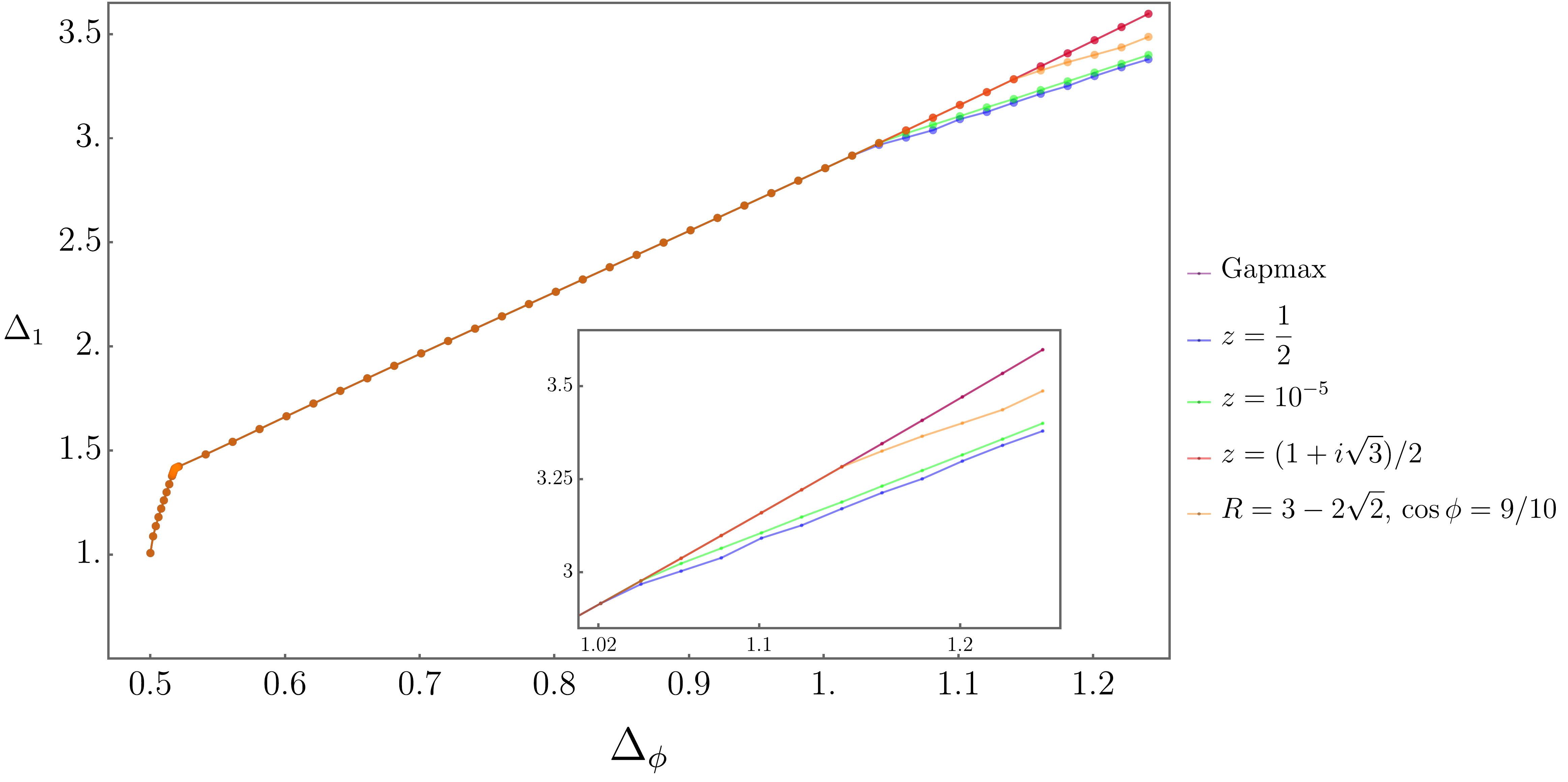}
	\caption{\label{comp}Leading scalar operator in the infimum correlator for different points on the Euclidean section as a function of $\Df$. Shown is also is an upper bound on the maximal scalar gap. All curves closely overlap until about $\Df\sim 1$ where the gap bound crosses $\Delta_{1}\sim d$.}		
\end{figure}

\begin{figure}[ht]
	\centering
	\includegraphics[width=12cm]{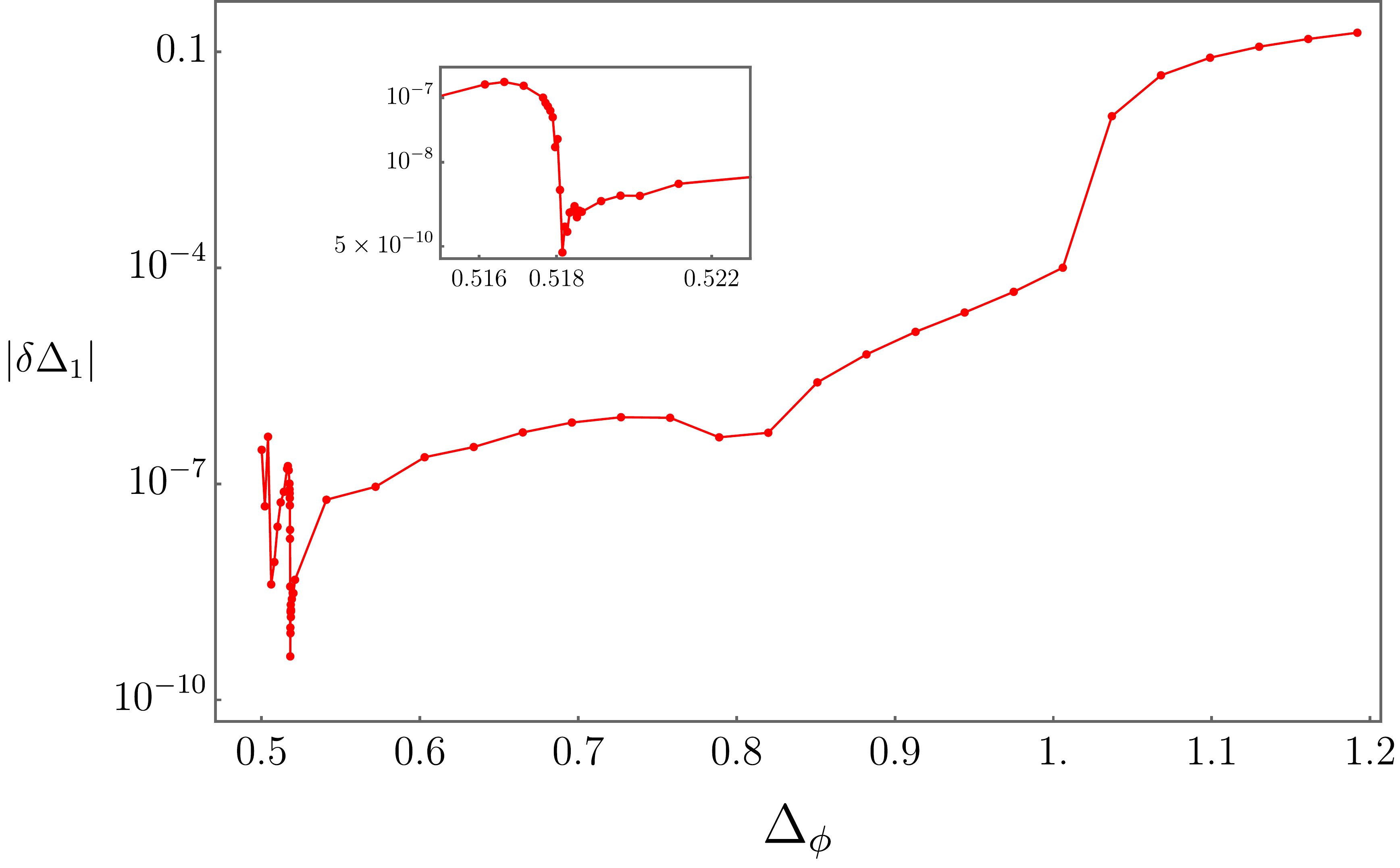}
	\caption{\label{diff} Difference in dimension of leading scalar operator between full correlator minimization and scalar channel correlator minimization.}
\end{figure}

We now take a closer look at the spectra of the minimizing correlators. Figure \ref{comp} show the dimension of the leading scalar operator in $\mathcal G_{\mbox{\tiny min};z,\bar z}$ as functions of $\Df$, depending on the choice of minimization point.
We find that to very high accuracy the minimization problem is saturated by the same correlator (i.e. the same solution to crossing) independently of the choice of point on the Euclidean section, as long as $\Df\lesssim 1$. That is, in this region the infimum correlator $\Gm$ is an actual correlator. Furthermore, in this region the full spectrum very closely matches the spectrum of the gap maximizing correlator, i.e.
\bea
\Gm\sim \mathcal G_{\mbox{\tiny gapmax}}\,, \qquad \Df\lesssim 1\,.
\eea
In particular it matches that of the critical 3d Ising model when $\Df= 0.518149$. Checking this to high precision is difficult for higher dimensional search spaces, since gap maximization requires a costly bissection procedure\footnote{An improved way to search for the boundary of the allowed region has been recently proposed \cite{Reehorst:2021ykw}.}, but to the extent we were able to do it we found very close agreement. As a proxy measure for how close, consider figure \ref{diff}, which shows the difference between spectra obtained by minimizing the full correlator or just minimizing contributions to the correlator from spin-0. The latter should be closer to the result of gap maximization in that channel. The figure reveals a very small difference for minimization at $z=\bar z=\frac 12$, which further shrinks as we make this value smaller.

The observation that minimization yields the same correlator almost independently of the choice of minimization point can be understood from our discussion in section \ref{sec:expect}: as long as the scalar gap is far from the spin-2 unitarity bound, the correlator is dominated for small values of $z,\bar z$ by the leading scalar, angle independent, contribution, which is minimized by maximizing the gap. However, as the dimension of the leading scalar approaches $3=d$, i.e. the unitarity bound in the spin-2 channel, we start to see an interplay between the contributions of the leading scalar and spin-2 operators, since now the value of the full correlator can be made smaller by pushing up the gaps in both these spin channels. Indeed, in this region the correlator is now approximately given by
\bea
(z\bar z)^\Df\mathcal G(\rho,\bar \rho)\sim 1+a_g R^{\Delta_g}+ a_T R^{\Delta_T} C_{\ell=2}^{(\frac{d-2}2)}(\cos \phi)+\ldots.
\eea
Minimization at different points $(z,\bar z)$ on the Euclidean section will lead to distinct minimal correlators $\mathcal G_{\mbox{\tiny min};z,\bar z}$ depending on the detailed contributions of each opeartor. 

\begin{figure}[ht]
	\centering
	\begin{tabular}{cc}
	\includegraphics[height=5cm]{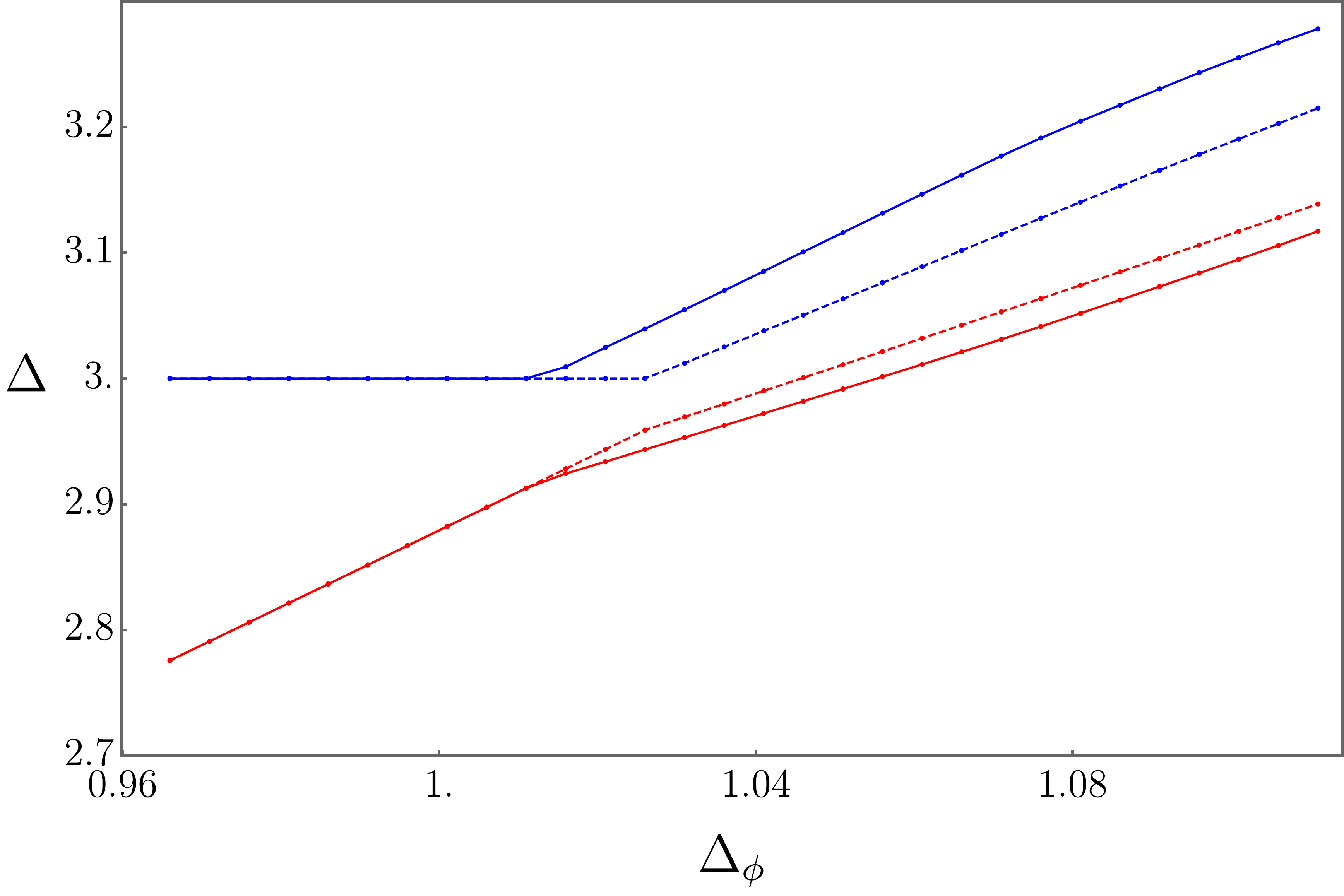}
	&
	\includegraphics[height=5cm]{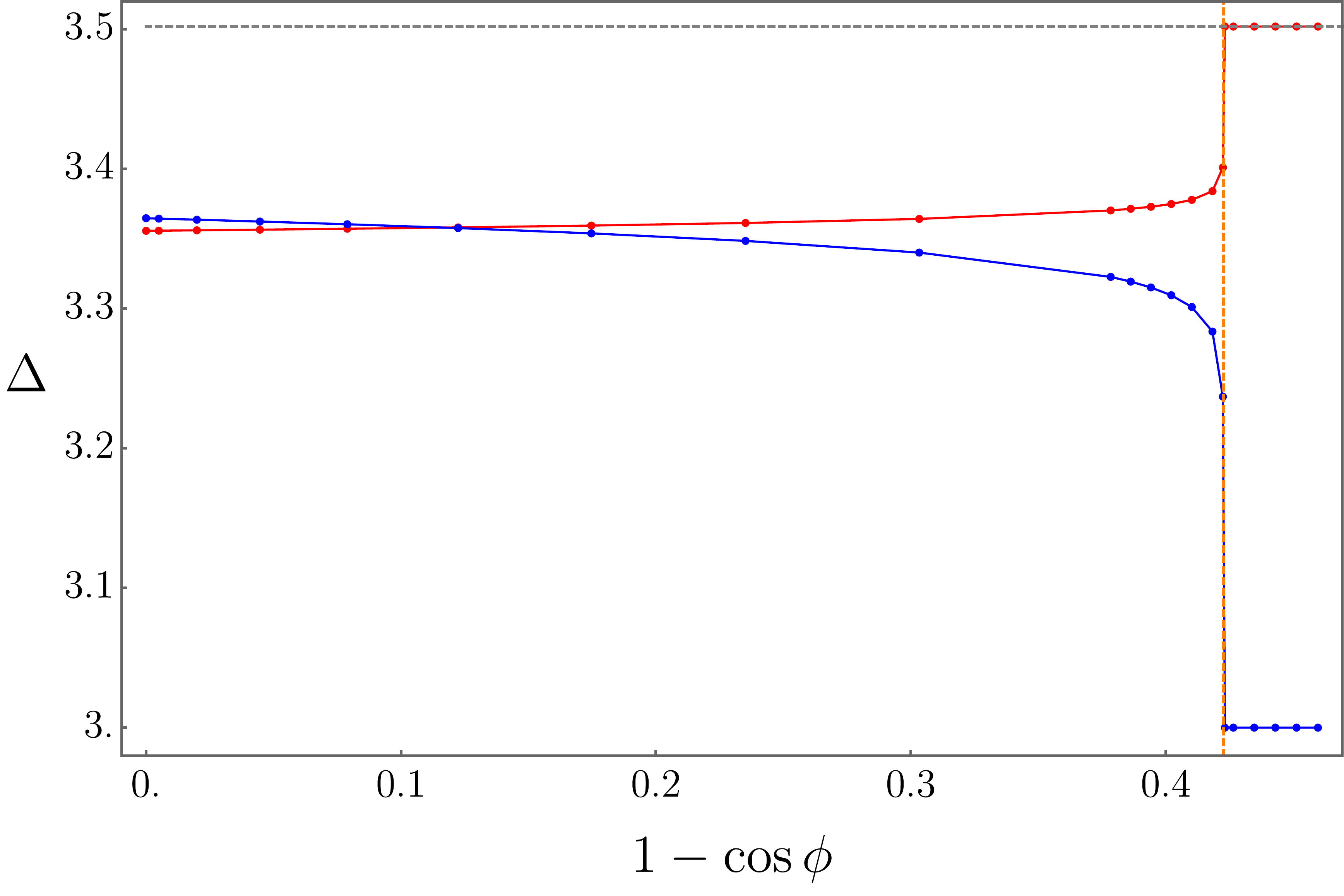}
	\end{tabular}
	\caption{\label{cross} Correlator spectra in the cross-over region. Left: dimensions of leading scalar and spin-2 operators, shown in red and blue respectively, in the spectrum of $\mathcal G_{\mbox{\tiny min},R}$. The solid lines correspond to spectra for  $R=3-2\sqrt{2}\sim 0.17$, and dashed lines for $R=5\times 10^{-3}$. Right: the same operators when minimizing for $R=3-2\sqrt{2}$ and different choices of the angle $\phi$. For $\cos\phi>1/\sqrt{3}$ the spin-2 Gegenbauer polynomial becomes negative and $T$ gets pushed down to unitarity.}
\end{figure}

We demonstrate this interplay in figure \ref{cross}. On the left, we show the leading operators in the scalar and spin-2 channel in the vicinity of the cross-over point where the gap bound hits $d$. For some $R$ dependent value of $\Df$, it becomes favourable to increase the gap in both sectors. In the limit where $R\to 0$ we expect the matching between gap maximization and correlator minimization to stop precisely for that value of $\Df$ where $\Delta_{\mbox{\tiny gapmax}}=d=3$. On the righthand figure we consider a specific $\Df>1$ (i.e. beyond the cross-over) and minimize the correlator for fixed $R$ but varying angle. As we move away from the forward limit, the scalar operator moves up until it hits the corresponding upper bound for that $\Df$. This can be understood from our simple approximation above, as increasing the angle makes the spin-2 contribution smaller. In fact, it can even become negative, which then favours pushing down the dimension of the $T$ operator to the unitarity bound.

\subsection{Maximization: no gap}
\label{sec:max}
\begin{figure}[ht]
	\begin{subfigure}{.5\textwidth}
		\centering
		\includegraphics[height=4.5cm]{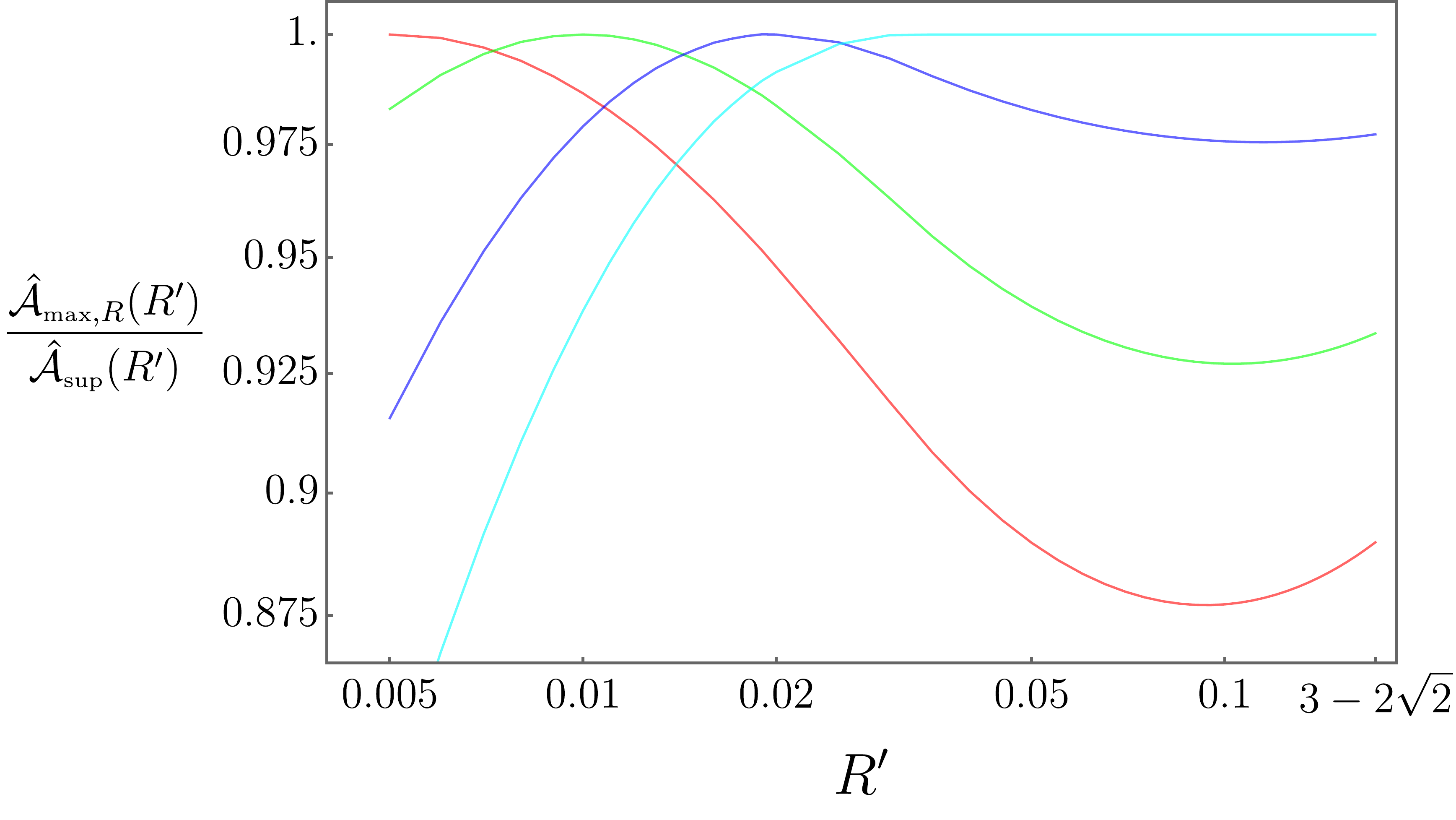}
			\end{subfigure}%
	\begin{subfigure}{.5\textwidth}
		\centering
		\includegraphics[height=4.5cm]{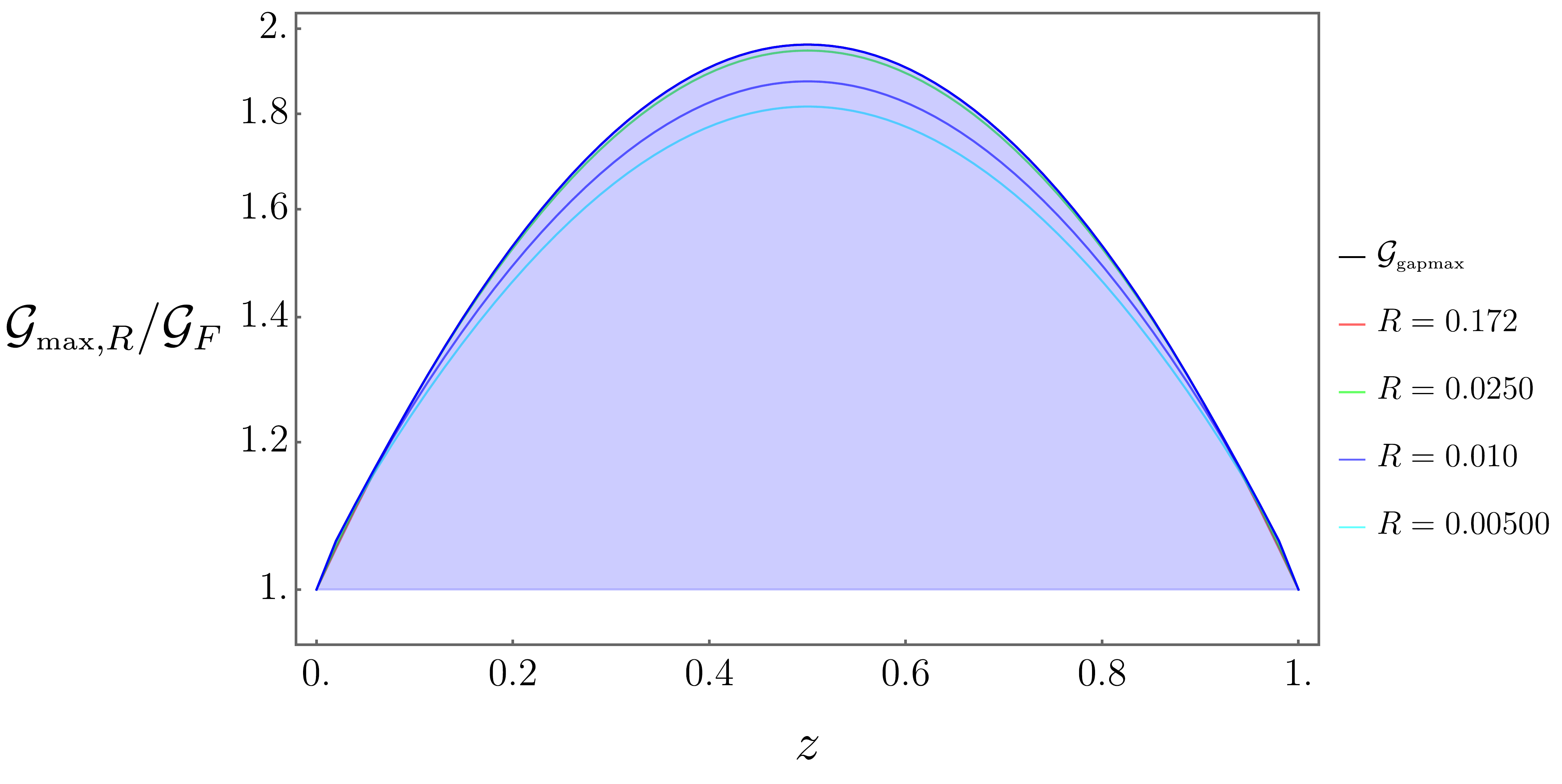}
	\end{subfigure}
	\caption{\label{envelope}Upper bound on correlator. On the right we plot $\mathcal G_{\mbox{\tiny max},R}(z)$ for various values of $R$. Their envelope defines $\GM(z)$.  This is seen more clearly in the lefthand figure.}
\end{figure}

Let us now turn to a closer look into the upper bounds on correlator values.
Figure \ref{envelope} shows the upper bound $\mathcal G_{\mbox{\tiny sup}}(z,\bar z)$ on the correlator along the line $z=\bar z$. The bound $\GM$ is the envelope of distinct maximizing correlators at distinct points, $ \mathcal G_{\mbox{\tiny max},z_i,\bar z_i}(z,\bar z)$. Hence the maximization problem generically leads to a continuous family of correlators labeled by the maximizing point.

\begin{figure}
    \centering
    \includegraphics[width=10cm]{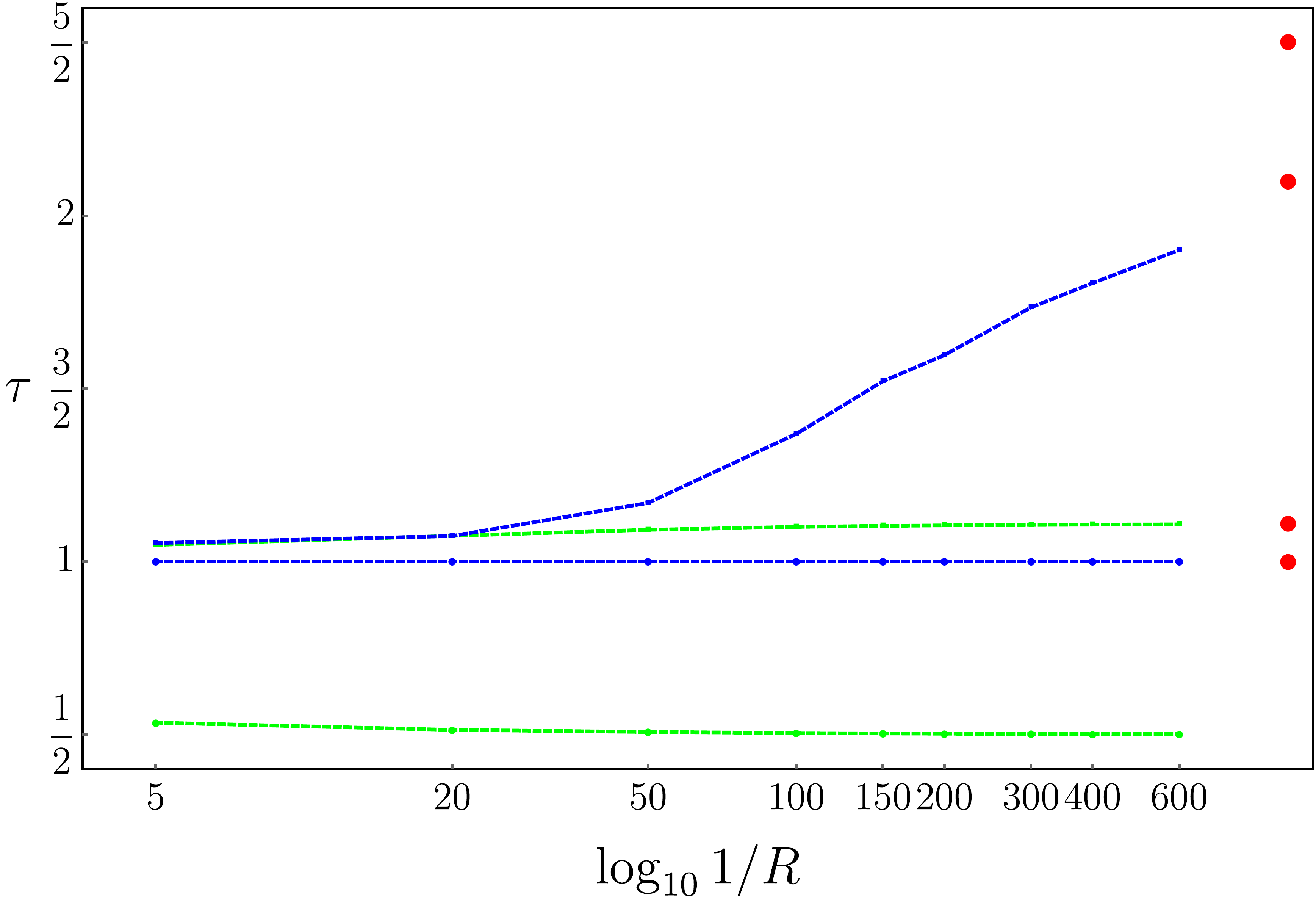}
    \caption{Twist spectrum of $\mathcal G_{R}$ as $R$ goes to zero. In green/blue $\ell=0/2$ operators. In the limit where $R\to 0$ the contribution of the operator at unitarity mimicks that of an operator with $\Delta=5/2$, and the full spectrum matches that of OPE maximization at shown as the red dots. See the main text for further details.}
    \label{fig:maxr}
\end{figure}

We can understand this by going back to the small $R$ approximation of the correlator, %
\bea
(z\bar z)^{\Df}\mathcal G(R)\sim 1+a_g R^{\Delta_g}+\ldots
\eea
where $\Delta_g$ is the dimension of the leading (scalar) operator, which we don't know a priori. Since we are not imposing any gap, maximizing the correlator at a fixed value of $R$ would tentatively place an operator at the lowest possible value of scaling dimension, namely the scalar unitarity bound, and maximize its OPE coefficient. However there is a snag, since a free field cannot appear in the OPE of two identical scalars. As we lower the scaling dimension towards unitarity, the conformal block for such an operator diverges, and this divergence must be cancelled by a vanishing OPE coefficient. In practice then we have a competition between wanting to lower the gap and keeping the OPE coefficient large. Writing $a_g\sim (\Delta_g-1/2)\hat a_g$ the correlator value at $R$ is maximized when
\ba
\partial_{\Delta_g} \left (a_g R^{\Delta_g}\right)=0 \Leftrightarrow \Delta_g-\frac 12 \sim \frac{1}{-\log(R)}
\ea
Hence we see that the maximizing correlator will indeed depend on the value of $R$. This is visible in figure \ref{fig:maxr} where we show the spectrum of the maximal correlator as a function of $R$.

Something very interesting happens in the limit where $R\to 0$. In this case the dimension of the leading operator in $\mathcal G_{\mbox{\tiny max},R}$ goes towards unitarity, and hence its OPE coefficient becomes vanishing, according to our reasoning above. This happens in such a way that the contribution of this operator still dominates the overall size of the correlator when we evaluate it close to the maximizing point. However, evaluating this correlator at some other value of the cross-ratio, i.e. $\mathcal G_{\mbox{\tiny max},R}(R')$ with $R\ll 1$ but $R'$ fixed, then the contribution of this operator will generically not be the dominant one. Since the corresponding OPE coefficient is going to zero we can use the identity
	\bea
	\lim_{\Delta\to \frac 12}(\Delta-\frac 12) G_{\Delta,0}(z,\bar z)\propto G_{5/2,0}(z,\bar z)\,.
	\eea
	and so the contribution of that operator becomes equivalent to that of an operator of dimension $5/2$.
	
\begin{figure}[ht]
    \centering
    \begin{tabular}{lr}
    \includegraphics[width=7.7cm]{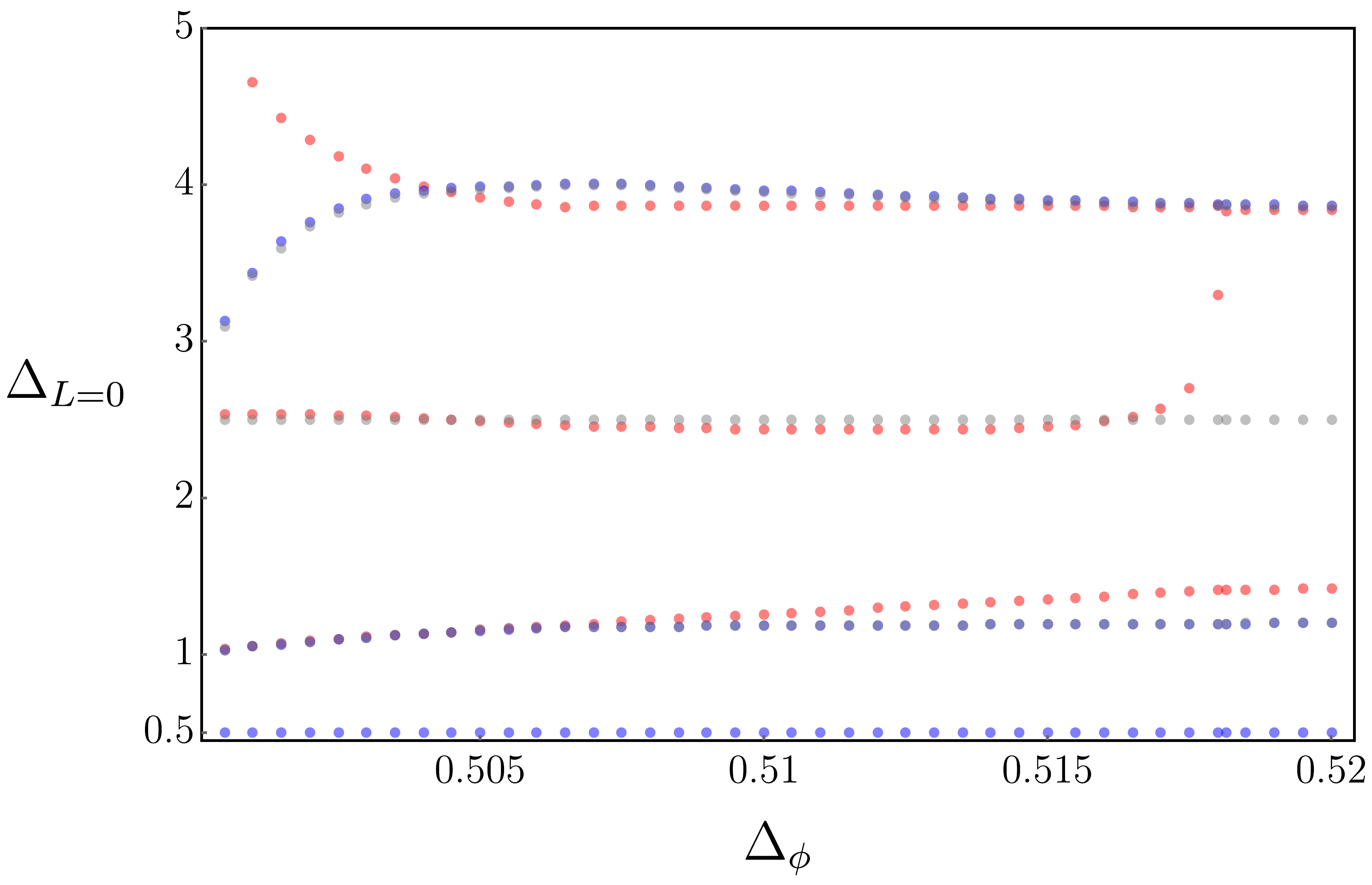}
    &
    \includegraphics[width=7.5cm]{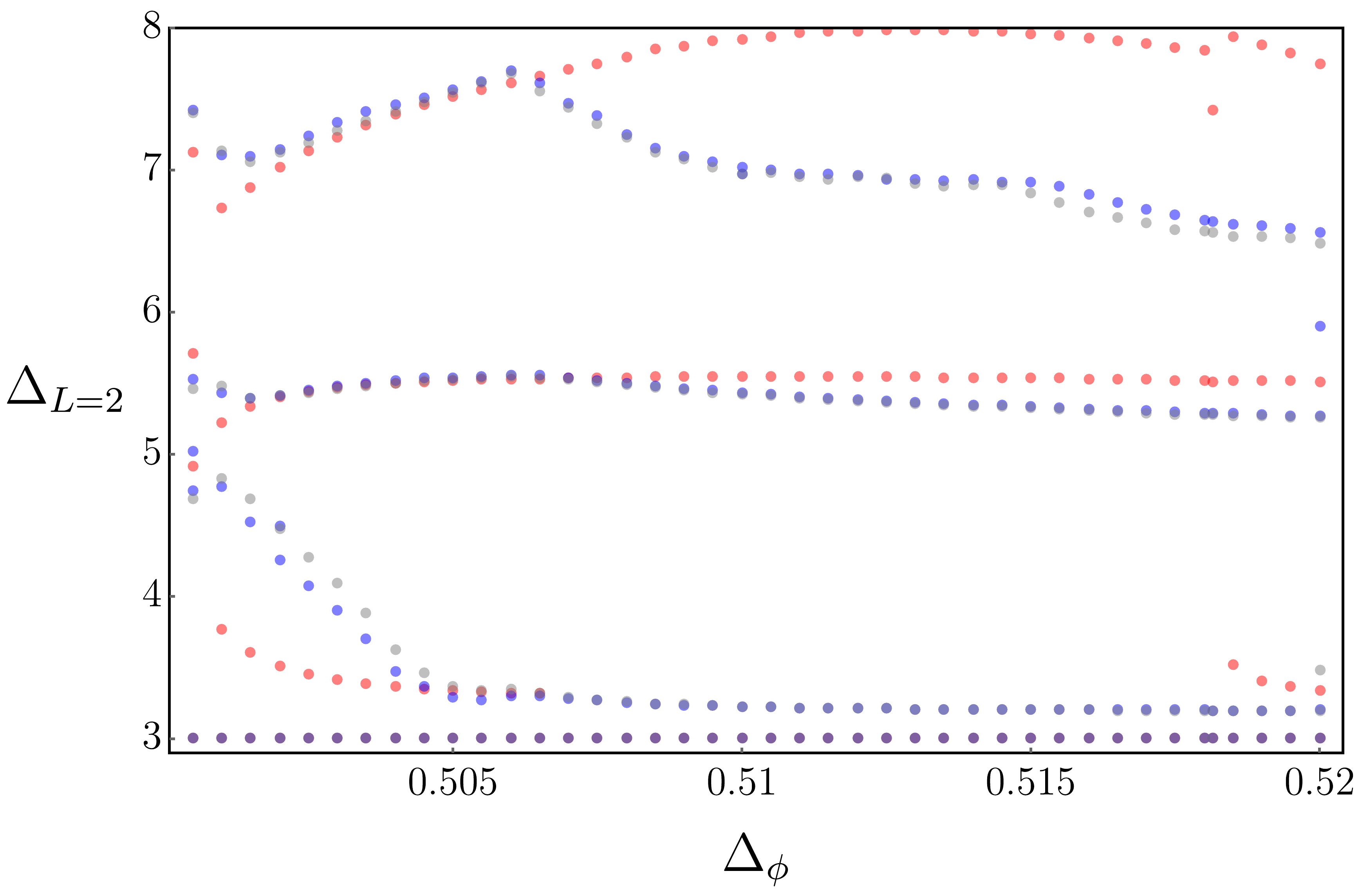}
    \end{tabular}
    \caption{Extremal correlator spectra. The blue dots correspond to correlator maximization at $R=10^{-600}$, the gray dots to OPE maximization at $\Delta=5/2$. The spectra closely match, once we identify the scalar at unitarity in the former to the $\Delta=5/2$ scalar in the latter. The red dots correspond to gap maximization. All three spectra closely match around $\Df \sim 0.505$.}
    \label{fig:opegapmax}
\end{figure}

	In other words, the family of maximal correlators labeled by $R$ tends to a universal correlator when $R\to 0$, although it does so non-uniformly as a function of the cross-ratio. Let us therefore set:
\ba
\mathcal G_{\mbox{\tiny max},0}(R')\equiv \lim_{R\to 0} \mathcal G_{\mbox{\tiny max},R}(R')\,.
\ea
	Let us comment on what we know about the properties of the spectrum of $\mathcal G_{\mbox{\tiny max},0}$. Firstly, the leading operator in $\mathcal G_{\mbox{\tiny max},0}(R')$ is now the ($R\to 0$ limit of) the subleading operator in the family $\mathcal G_{\mbox{\tiny max},R}$, and it will have some scaling dimension which we denote $\Delta_g^0$. Secondly, the correlator will contain a scalar operator of dimension $5/2$. Furthermore, because of the way in which this operator arose, we expect that its OPE coefficient should be quite large. In fact, it is reasonable to expect that it should be as large as possible, so that $\mathcal G_{\mbox{\tiny max},0}$ should actually be closely related to that correlator which maximizes the OPE coefficient of the operator at $\Delta=5/2$ {\em without assumptions on the gap}:\footnote{In the notation of section \ref{sec:corrspace}, this correlator is given by:
	\bea
	\mathcal G_{\mbox{\tiny opemax},\Delta=5/2}=\underset{\mathcal G \in \mathfrak G}{\mbox{arg max}}\quad a_{5/2,0}\,.
	\eea
	}
	\bea
	\mathcal G_{\mbox{\tiny max},0}(z,\bar z)\overset{?}{=} \mathcal G_{\mbox{\tiny opemax},\Delta=5/2}
	\eea
We compare the spectrum of $\mathcal G_{\mbox{\tiny max},0}$ against that of $\mathcal G_{\mbox{\tiny opemax},\Delta=5/2}$ in figure \ref{fig:opegapmax} (see also figure \ref{fig:maxr}). We can see that up to the mapping between the two scalar operators at unitarity and $\Delta=5/2$, the spectra indeed closely match.

However, we also find a surprise. In the same plot we show the spectrum of the gap maximization correlator $\mathcal G_{\mbox{\tiny gapmax}}$. Recall that we had found that this closely matches that of correlator {\em minimization}. Remarkably, it appears that around some critical value $\Df\sim 0.505$, the three correlators actually match! In fact, the subleading scalar operator in $\mathcal G_{\mbox{\tiny gapmax}}$, which corresponds to $\epsilon'$ in the 3d Ising model with $\Delta\sim 3.84$, comes down with decreasing $\Df$ and becomes the $5/2$ operator. This suggests that at this point could sit an interesting CFT, one which would be characterized by the existence of a protected operator of dimension $5/2$. This CFT is singled out by simultaneously satisfying an OPE maximization bound on this operator and saturating a bound on the maximal gap. It would be interesting to extend our analysis of this theory further by considering a multiple correlator setup.

%\begin{figure}[ht]
%	\centering
%	\includegraphics[scale=0.4]{Figures/max20.pdf}
%	\caption{\label{max20}The first two operators for the extremal solution that maximize the correlator at $z=\bar{z}=1/2$. There is a hole that the operator with dimension around 0.7 disappear.}
%\end{figure} 

\subsection{Maximization: with a gap}
Let us now discuss correlator maximization assuming a gap in the spectrum of scalar operators. 
\begin{figure}[ht]
	\centering
	\includegraphics[width=15cm]{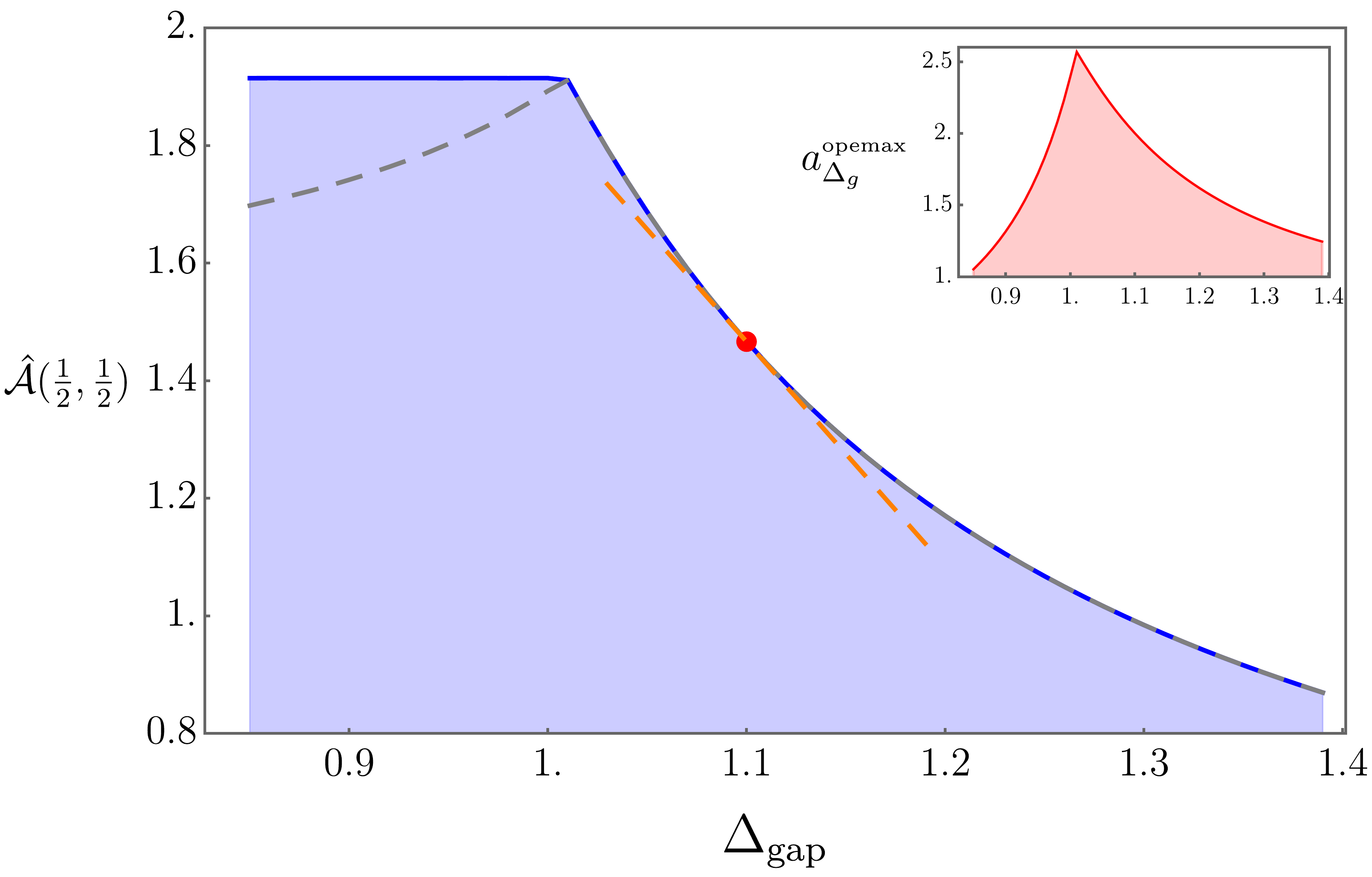}
	\caption{\label{max20} Correlator maximization at the crossing symmetric point as a function of the gap. The resulting correlator matches the OPE maximization correlator (shown as a gray dashed line) above a certain critical gap. Both correlators agree with the generalized free boson at a gap $\Delta_g=2\Df$ shown by the red dot. The dashed orange line which is tangent to the bound curve at the GFF point represents the leading correction to the correlator from an AdS$_4$ contact interaction. The inset shows a bound on the OPE coefficient $a_{\Delta_g}^{\mbox{\tiny opemax}}$ as a function on the gap. It shows a maximum at the point below which correlator maximization and OPE maximization give different results.}
\end{figure} 

In figure \ref{max20} we show the upper bound for the value of the correlator at the crossing symmetric point as a function of the gap at a specific value of $\Df$. Other choices lead to similar looking figures. For large enough gaps, we find that the maximizing correlator is independent of the precise maximizing point, so that the supremum correlator $\GM$ is physical. Indeed, in this case the bound is very closely saturated by the correlator saturating OPE maximization with the same gap. There are minute differences between the two, which seemingly get stamped out as we decrease $r$, similarly to what we saw for gapmax versus correlator minimization.  A particularly interesting point is where we set the gap to be equal to $2\Df$. In this case, the 1d functionals discussed in section \ref{sec:exact} imply that the optimal OPE and correlator maximization bounds should not only match, but should correspond to the generalized free boson correlator. Our numerical results agree with this and we've checked the result is unchanged by moving away from the line $z=\bar z$.

As we lower the gap, we eventually reach a critical value $\Delta_g^c$, which is approximately equal to one in a wide range of $\Df$, where the OPE max correlator $\mathcal G_{\mbox{\tiny opemax}}$ and $\GM$ stop matching. Also, if we lower the gap further, the value of the correlator varies very little. This can be understood by a closer examination of the OPE maximization problem. Indeed, as shown in the inset of the figure, the maximal value of the OPE coefficient as a function of the gap, peaks at about $\Delta_g\sim \Delta_g^c$. So, for gaps lower than this, the maximizing correlator can now benefit from a tradeoff between having an operator at a dimension larger than the gap but with larger OPE coefficient.

The region where both OPE and correlator maximization agree is very interesting. Zooming in to the vicinity of the generalized free field point, we have verified that the maximal solution is well approximated
by the boundary correlator of a weakly coupled field $\Phi$ in AdS$_4$ with a $\Phi^4$ contact interaction:
\bea
\mathcal G_{\mbox{\tiny sup}}(z,\bar z| \Delta_g) \underset{\Delta_g\sim 2\Df}\sim \mathcal G^B(z,\bar z)+\frac{(\Delta_g-2\Df)}2\, D_{\Df}(z,\bar z)
\eea
where the crossing symmetric $D$ function is the overlap of four bulk-to-boundary propagators in AdS \cite{Freedman:1998bj,DHoker:1999mic,Hijano:2015zsa}, with conformal block expansion
\bea
D_{\Df}(z,\bar z)=\sum_{n=0}^\infty\left[ a_{n}^{(1)} G_{\Delta_n,0}(z,\bar z)+a_{n,0}^{\mbox{\tiny gff}} \gamma_{n,0} \partial_{\Delta} G_{\Delta_n,0}(z,\bar z)\right]
\eea
where

\ba
    a_n^{(1)}&=\frac 12 \partial_n \left(a_{n,0}^{\mbox{\tiny gff}} \gamma_{n,0}\right)\\
   a_{n,0}^{\mbox{\tiny gff}} \gamma_{n,0} &=
   \frac{2^{1-2 n} \Gamma \left(\Df+\frac{1}{2}\right) \Gamma (\Df+n)^3 \Gamma \left(2 \Df-\frac{d}{2}+n\right)^2}{(n!)^2\Gamma
   (\Df)^3 \Gamma \left(2 \Df-\frac{d}{2}\right) \Gamma \left(\Df+n+\frac{1}{2}\right) \Gamma \left(2
   \Df-\frac{d}{2}+2 n\right)}
\ea
For instance we have verified that close to the GFF point the spectrum of the solution is well described by the GFF one plus corrections to the scalar operators controlled by $\gamma_{n,0}$.
This suggests that the line of solutions running from $\Delta_g^c$ and $\Delta_{\mbox{\tiny gapmax}}$ is related to a scalar theory in AdS$_4$, whose dynamics at small coupling is described by a simple contact interaction.
A similar line of solutions had been observed in $d=1$ CFTs where in fact it was checked the agreement extended to $O(g^3)$ with $g=\Delta_g-2\Df$ the effective coupling. It would be interesting to investigate what happens here where the contact interaction in AdS$_4$ is classically marginal. Increasing the coupling, making the interaction more attractive, lowers the energy of states and the gap goes down. However there seems to a limit for doing this, since when we hit $\Delta_g=\Delta_g^c$ we must move on to a very different branch of solutions; for instance we find that the stress-tensor becomes a part of the spectrum for gaps lower than $\Delta_g^c$. Conversely making the interaction repulsive increases the energy. We cannot do this indefinitely however: at some point we reach the maximal allowed value of the gap. This is interesting: were we to sit at $\Df=\Delta_{\sigma}^{\mbox{\tiny Ising}}$ (say by tuning the AdS mass), we know that the maximal gap is achieved by the 3d Ising correlator, which is also described by a scalar field with a quartic interaction, but of course on the {\em boundary} of AdS. The transition from the family of theories labeled by the gap to the Ising solution is therefore most likely discontinuous, since the latter is local (it contains a stress tensor), whereas our family does not (as we've checked). This is what happens in the $d=1$ case where the gap max solution is a free fermion.\footnote{In that case there is an order of limits issue: for any solution with $\Delta_g$ strictly smaller than $\Delta_{\mbox{\tiny gapmax}}$, its spectrum is always different from that of the gap max solution at sufficiently high energies.} It would be very interesting to understand this family of solutions better, and how it links with yet another family of solutions linking 3d Ising and generalized free fields, namely the long-range Ising family \cite{Paulos:2015jfa,Behan:2017dwr,Behan:2017emf,Behan:2018hfx}.

\begin{figure}[t!]
	\begin{subfigure}{.5\textwidth}
		\centering
		\includegraphics[width=.9\linewidth]{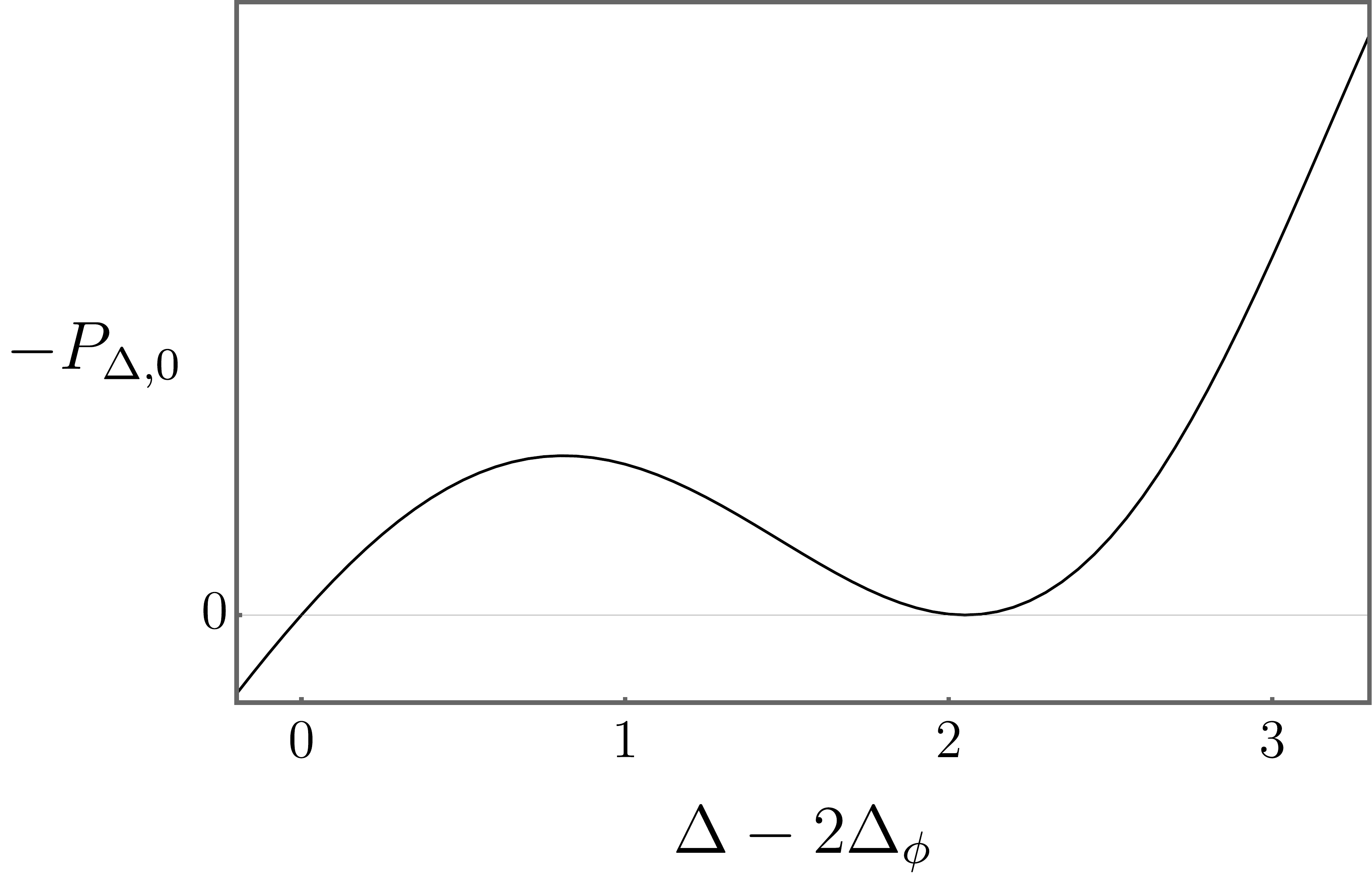}
	\end{subfigure}%
	\begin{subfigure}{.5\textwidth}
		\centering
		\includegraphics[width=.9\linewidth]{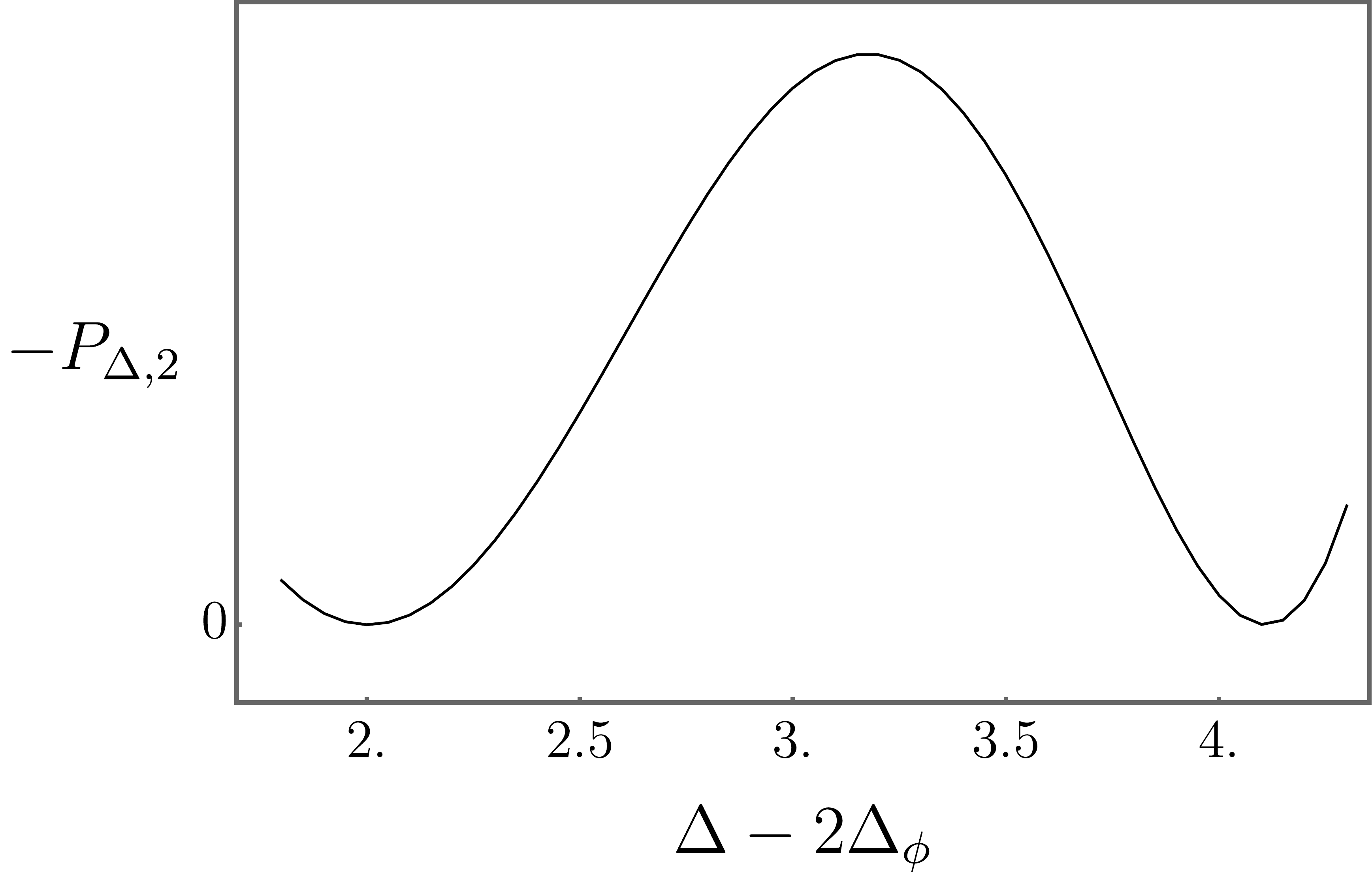}
	\end{subfigure}
	\caption{\label{functional}Polyakov blocks from numerical results. They are defined by $-P_{\Delta,\ell}(z,\bar z):= \Omega_{z,\bar z}(\Delta,\ell)-G_{\Delta,\ell}(z,\bar z |\Df)$. Here we evaluate them in the case where $\Omega$ is the functional establishing an upper bound on th correlator at the crossing symmetric point and in the presence of a gap $\Delta_g\geq 2\Df$. The spin-2 Polyakov block is non-negative for all $\Delta$ above the unitarity bound.}
\end{figure}

\section{Discussion}
In this paper we have initiated an exploration of the CFT landscape from a new perspective: that of optimization of correlator values. We have found that in some circumstances, this new viewpoint coincides or closely matches with more familiar ones of OPE maximization or gap maximization, in others it leads to new extremal solutions that would be difficult to find otherwise. An example is the gap maximizing correlator around $\Df=0.505$ which has an operator of dimension $\Delta=5/2$. It would be very interesting to try to isolate this theory as an island in a multiple correlator setup and to try to understand theoretically how such a protected operator can arise. Another example is the fact that the GFF solution and its nearby deformations in AdS saturate an OPE/correlator maximization bound. This immediately implies that in the vicinity of $\Df=d/4$ the long-range Ising model family also saturates this bound. It is likely this fact can be used to isolate the theory. We hope to report on this in the near future.

There are several obvious generalizations of our work, such as different spacetime dimensionality, including global symmetries and so forth. It could also be interesting to consider different optimization problems, where instead of optimizing over correlator values we would optimize over value variations, e.g. derivatives of the correlator. There is some motivation for doing this from the analogy with the S-matrix bootstrap, where this would correspond to optimizing over couplings to resonances. It might also be interesting to consider non-linear optimization targets in the multiple correlator bootstrap. For instance, is the determinant of the matrix of correlators involving $\sigma$ and $\epsilon$ for the 3d Ising model the minimal possible for all CFTs?

We have obtained bounds by constructing functionals numerically. These functionals constitute approximations to master functionals \cite{Caron-Huot:2020adz,Paulos:2020zxx} which encode higher dimensional versions of Polyakov blocks \cite{Sen:2015doa,Gopakumar:2016wkt,Gopakumar:2016cpb,Gopakumar:2018xqi,Mazac:2018ycv}. We can plot them numerically (they are simply the functional actions) and see that they have nice positivity properties. In particular, for the maximization problem with gap $2\Df$, where the blocks have GFF spectra, they cannot and do not match those of the Polyakov-Regge expansion of the works \cite{Mazac:2019shk,Caron-Huot:2020adz}. For instance our functionals do not have any nice structure in the odd-spin sector, and they are positive below twist $2\Df$ for all $\ell>0$. It would be very interesting to investigate if instead they could be matched to the fully crossing-symmetric Polyakov blocks defined in \cite{Gopakumar:2021dvg}.

\section*{Acknowledgments}
We would like to thank Simons Collaboration on the Nonperturbative Bootstrap for providing opportunities for discussion and collaboration while this work was being conducted.
\pagebreak

\appendix
\section{Numerical parameters}\label{detail}

All computations done in this work used the the {\tt JuliBootS} package \cite{Paulos:2014vya}. The table below specifies the parameters in our numerical calculations. When $\mathrm{mmax}=1$, nmax is equivalent to the usual $\Lambda$ used in SDPB \cite{Simmons-Duffin:2015qma} by $\Lambda=2\mathrm{nmax}+1$. For almost all the plots, we used using the parameters No.1. The Fig~\ref{fig:maxr} and Fig~\ref{fig:opegapmax} correspond parameters No.2. Fig~\ref{cross} left and right correspond to parameters No.3 and No.4. The Fig~\ref{comp} and Fig~\ref{max20} correspond to parameters No.5. 

\begin{table}[h!]
\centering

\begin{tabular}{SSSSS} \toprule
{No.} & {nmax} & {mmax} &  {Spins}  \\ \midrule
1 &    11  & 1 &  \(\{0,2, ..., 28 \} \cup\{46,48,50\}\) \\
2 &    20  & 1 &  \(\{0,2, ..., 40 \} \cup\{64,66,68,70\}\) \\
3 &    10  & 1 &  \(\{0,2, ..., 40 \} \cup\{64,66,68,70\}\) \\ 
4 &    10  & 1 & \(\{0,2, ..., 28 \} \cup\{46,48,50\}\) \\
5 &    13  & 1 &  \(\{0,2, ..., 40 \} \cup\{58,60,76,78,96,98,100\}\)   \\
\bottomrule
\end{tabular}   
\caption{\label{Parameter}Parameters used in numerical computations.}
\end{table}

\small
\parskip=-10pt
\bibliography{mybib}
\bibliographystyle{jhep}

\end{document}